\documentclass[11pt,draftclsnofoot,onecolumn]{IEEEtran}

\usepackage{amsfonts}
\usepackage{times}
\usepackage{latexsym}
\usepackage{amssymb}
\usepackage{amsmath}
\usepackage{cite}
\usepackage{verbatim}

\newcommand{\bydef}{\triangleq}

\def\SNR{{\textsf{SNR}}}

\def\bydef{:=}

\def\bb0{{\mathbb{0}}}

\def\bydef{:=}

\def\bb{{\mathbf{b}}}
\def\bc{{\mathbf{c}}}

\def\bee{{\mathbf{e}}}

\def\bg{{\mathbf{g}}}
\def\bh{{\mathbf{h}}}

\def\bn{{\mathbf{n}}}

\def\br{{\mathbf{r}}}

\def\bt{{\mathbf{t}}}

\def\bw{{\mathbf{w}}}
\def\bx{{\mathbf{x}}}
\def\by{{\mathbf{y}}}
\def\bz{{\mathbf{z}}}
\def\b0{{\mathbf{0}}}

\def\bA{{\mathbf{A}}}
\def\bB{{\mathbf{B}}}
\def\bC{{\mathbf{C}}}
\def\bD{{\mathbf{D}}}

\def\bG{{\mathbf{G}}}
\def\bH{{\mathbf{H}}}
\def\bI{{\mathbf{I}}}

\def\bQ{{\mathbf{Q}}}

\def\bU{{\mathbf{U}}}
\def\bV{{\mathbf{V}}}
\def\bW{{\mathbf{W}}}

\def\bbC{{\mathbb{C}}}

\def\bbE{{\mathbb{E}}}

\def\bbZ{{\mathbb{Z}}}

\def\bydef{:=}

\def\sf0{{\mathsf{0}}}

\newcommand{\expeq}{\stackrel{.}{=}}

\newcommand{\expl}{\stackrel{.}{\le}}

\usepackage{graphicx}
\usepackage{amssymb}
\usepackage{amsfonts}
\usepackage{amsmath}
\begin{document}
\newtheorem{thm}{Theorem}
\newtheorem{lemma}{Lemma}
\newtheorem{rem}{Remark}
\def\proof{\noindent\hspace{0em}{\itshape Proof: }}
\def\endproof{\hspace*{\fill}~\QED\par\endtrivlist\unskip}
\def\mapright#1{\smash{\mathop{\le}\limits_{#1}}}
\def\mapequal#1{\smash{\mathop{=}\limits_{#1}}}
\title{On the Capacity and Diversity-Multiplexing Tradeoff of the Two-Way Relay Channel}
\author{Rahul~Vaze and Robert W. Heath Jr. \\
The University of Texas at Austin \\
Department of Electrical and Computer Engineering \\
Wireless Networking and Communications Group \\
1 University Station C0803\\
Austin, TX 78712-0240\\
email: vaze@ece.utexas.edu, rheath@ece.utexas.edu
\thanks{This work was funded by DARPA through IT-MANET grant no. W911NF-07-1-0028.}}

\date{}
\maketitle
%%\footnotetext[1]{}
\noindent
\begin{abstract}
This paper considers a multiple input multiple output (MIMO) two-way relay channel, where two nodes want to exchange data with each other using multiple
relays. An iterative algorithm is proposed to achieve the optimal achievable rate region, when each relay employs an amplify and forward (AF) strategy.
 The iterative algorithm solves a power minimization problem at every step, subject to minimum signal-to-interference-and-noise ratio constraints, 
which is non-convex, however, for which the Karush Kuhn Tuker conditions are sufficient for optimality. The optimal AF strategy assumes
global channel state information (CSI) at each relay. To simplify
the CSI requirements, a simple amplify and forward strategy, called
dual channel matching, is also proposed, that requires only local
channel state information, and whose achievable rate region is close
to that of the optimal AF strategy.
In the asymptotic regime of large number of relays, we show that
the achievable rate region of the dual channel matching
and an upper bound differ by only a constant term and
establish the capacity scaling law of the two-way relay channel.
Relay strategies achieving optimal diversity-multiplexing tradeoff are also
considered with a single relay node. A compress and
forward strategy is shown to be optimal for achieving diversity multiplexing tradeoff for the full-duplex case, in
general, and for the half-duplex case in some cases.
\end{abstract}
\section{Introduction}
\begin{figure}
\centering
\includegraphics[height= 1in]{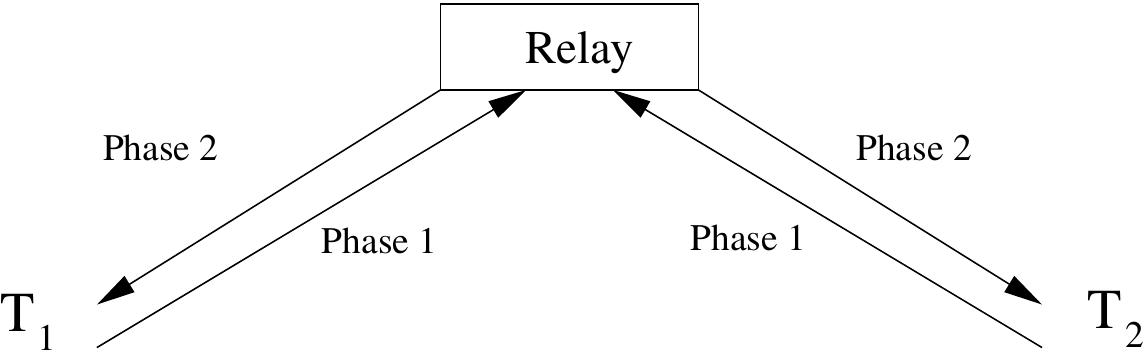}
\caption{Two way relay channel communication protocol}
\label{twowayarmin}
\end{figure}

We consider a multiple antenna two-way relay channel as shown in Fig. \ref{twowayarmin}, where two nodes $T_1$ and $T_2$ want to
exchange information with each other with the help of a relay node and all the
nodes are equipped with one or more than one antenna.
The two-way relay channel models the communication scenario where the
destination terminal also has some data to send to source terminal
e.g. downlink and uplink in cellular communication, or packet acknowledgments
in a wireless network.
The general discrete memoryless two-way relay channel was introduced in
\cite{Muelen1971}, and the multiple antenna two-way relay channel in \cite{Rankov2005}. In the literature, the two-way relay channel
is also known by several other names, including the:
bidirectional relay channel \cite{Kim2007,Kim2008,Boche2007}
and analog network coding \cite{Dina2006}.

A specific embodiment of a multiple antenna two-way relay channel  that
assumes half-duplex relays and
the absence of a direct path between source and destination was proposed in
\cite{Rankov2005}. An illustration is provided in Fig. \ref{twowayarmin}.
As shown in Fig. \ref{twowayarmin}, in phase $1$ or the first time slot,
both terminals $T_1$ and $T_2$ are scheduled to transmit
simultaneously while the relay receives. In phase $2$ or the second time slot,
the relay is scheduled to transmit while terminals
$T_1$ and $T_2$ receive. The key idea with the two-way relay channel is that each terminal
can cancel the interference (generated by its own transmission) from the
signal it receives from the relay to recover the transmission from the other terminal.
The idea is reminiscent of work in network coding \cite{Yeung2003},
though note that here the coding is done in the analog domain,
 \cite{Dina2006} rather than in digital domain \cite{Yeung2003}.
In this paper we only consider multiple antenna two-way
relay channel and for brevity, drop the prefix multiple antenna from here
onwards.

There has been a growing interest in finding the capacity region of
the two-way relay channel with a single relay node \cite{Rankov2006,
Kim2007,Kim2008, Boche2007, Tse2008,Nazer2007,Narayan2007,
Popovski2007, Shengli2008}. Achievable sum rate expressions (sum of
the rates achievable from $T_1 \rightarrow T_2$ and $T_2 \rightarrow
T_1$ links) have been derived in \cite{Rankov2005} and
\cite{Boche2007,Kim2007,Kim2008}, for the half-duplex two-way relay
channel, using amplify and forward (AF), decode and forward (DF) and
compress and forward (CF) at the relay. It is shown that in a
two-way relay channel, it is possible to remove the $\frac{1}{2}$
rate loss factor in spectral efficiency due to the half duplex
assumption on the nodes. For a general full-duplex two-way relay
channel with a single relay node ($T_1$, $T_2$ and relay can
transmit and receive at the same time)
 achievable rate regions are derived in
\cite{Rankov2006} for AF, DF, and CF. For the AWGN two-way relay channel
(no fading), using nested lattice coding and DF at the relay, the
achievable rate region has been shown to be very close to the upper bound
for all SNRs \cite{Nazer2007,Narayan2007}. Using the deterministic
channel approach, the achievable rate region has been shown to be at most
three bits away from the upper bound for the full-duplex two-way relay channel
\cite{Tse2008}. The capacity region of the two-way relay channel
has also been studied in \cite{Popovski2007, Shengli2008}, where
in \cite{Shengli2008}, it has been shown that in
the low SNR regime the upper bound can be achieved by choosing a suitable
relay mapping function, together with LDPC codes.
The achievable rate region \cite{Kim2007,Kim2008, Tse2008,Nazer2007,Narayan2007,Boche2007,Rankov2006,Popovski2007, Shengli2008} does not meet the upper bound \cite{Kramer2003}, in
general. Consequently, the problem of finding the
capacity region of the two-way relay channel is currently open.

The problem of finding the capacity region of the two-way relay channel
becomes even more challenging when there are multiple relay nodes that can help $T_1$ and $T_2$, and to the best of our knowledge 
has not been addressed in the literature.
The problem becomes hard, because it is known that for the one-way relay channel
with multiple relay nodes, DF does not work well \cite{Kramer2005},
while the partial DF and distributed CF \cite{Kramer2005} lead to complicated
achievable rate regions that are very hard to compute and analyze. The same
conclusion holds true for the two-way relay channel; the
only simple strategy that is well suited for multiple relay nodes
is AF. With this motivation, in this paper we attempt to find the optimal
relay beamformers that maximize the achievable rate region of the two-way
relay channel with AF.
For the one-way relay channel with multiple relays,
optimal relay beamformers have been found \cite{Yi2007}, however,
they are not known for the two-way relay channel.

For the case when both $T_1$ and $T_2$ have a single antenna, and
each relay has an arbitrary number of antennas, we solve the problem of
finding optimal relay beamformers by recasting it
as an iterative power minimization algorithm.
The iterative algorithm, at each step, solves a power minimization
problem with minimum signal-to-interference-noise
(SINR) constraints, for which satisfying the
Karush Kuhn Tucker (KKT) conditions \cite{Boyd2004,Wiesel2006} are sufficient for
optimality.
We consider both the sum power constraint across relays, as well as
an individual relay power constraint. The optimal AF solution
requires each relay to have channel state
information (CSI) for all relays and leads to an achievable rate region that cannot be expressed in closed form.

For the case when each relay knows its own CSI, finding the optimal AF strategy is quite hard and intractable, even for
the one-way relay channel case \cite{Yi2007}. To remove the global CSI requirement, and
to obtain a simple achievable rate region expression, next, we
propose a simple AF strategy, called dual channel matching strategy, which works for any number of antennas at $T_1$ and $T_2$. In dual channel matching,
relay $k$ transmits the received signal multiplied
 with  $(\bG_k^{*}\bH_k^{*} + \bH_k^{r*}\bG_k^{r*})$, if the
channel between $T_1$ and relay $k$ is $\bH_k$, between relay $k$ and $T_2$ is
 $\bG_k^{r}$, between $T_2$ and relay $k$ is $\bG_k$ and between relay $k$ and $T_1$ is $\bH_k^r$. Using dual channel matching, we lower bound the
achievable rate region of the optimal AF strategy, which is unknown for more than one antenna at $T_1$ and $T_2$, and bound
the gap between the optimal AF strategy and the upper bound.
The dual channel matching is quite simple to implement
and its achievable rate region can be shown to be quite close
(by simulation) to the optimal AF strategy, when $T_1$ and $T_2$ each have
single antenna.

We upper bound the capacity region of the two-way relay channel
using the cut-set bound \cite{Cover2004} on the broadcast cut
$T_1$ ($T_2$), and $r_1, r_2, \ldots, r_K$, and the multiple access cut
 $r_1, r_2, \ldots, r_K$ and $T_2$ ($T_1$), over all possible
two phase protocols (with different time allocation between first and
second phase).
We show that the gap between the upper and lower bound
(dual channel matching) is quite small for small values of $K$. In
the limit $K \rightarrow \infty$, we show that the gap is constant with
increasing $K$, and thus establish the scaling law \cite{Bolcskei2006} of the capacity region of the two-way relay channel,
which shows that $\frac{M}{2}\log K$ bits can be transmitted from both
$T_1\rightarrow T_2$ and $T_2\rightarrow T_1$, simultaneously.

We also consider the problem of finding relay transmission
strategies to achieve
the optimal diversity multiplexing (DM)-tradeoff \cite{Zheng2003}
of the two-way relay channel with a single relay node, in the presence of
a direct path between $T_1$ and $T_2$. The DM-tradeoff captures
the maximum rate of fall of error probability with signal to noise ratio
($\SNR$), when rate of
transmission is increased as $r\log \SNR$. The DM-tradeoff for the two-way
relay channel is a two-dimensional region spanned by the
$\left(d_{12}(r_{12},r_{21}),d_{21}(r_{12},r_{21})\right)$,
where $d_{12}$ and $d_{21}$ are the negatives of the exponent of the
probability of error from $T_1\rightarrow T_2$ and $T_2\rightarrow T_1$,
respectively, when
$T_1$ is transmitting at rate $r_{12}\log \SNR$ and $T_2$ at $r_{21}\log \SNR$.
The DM-tradeoff for the one-way
relay channel has been studied in \cite{Azarian2005,Belfiore2007,EliaDec72005,Yuksel2007}, where notably in \cite{Yuksel2007}, it has
been shown that the CF strategy achieves the DM-tradeoff for both the full-duplex
as well as the half-duplex case. The DM-tradeoff of the two-way relay channel
has been recently studied in \cite{Mitran2008},
 where upper and lower bounds are obtained on the DM-tradeoff
 which are shown to match for the case when each node has
a single antenna.

We first consider the full-duplex two-way relay channel and
show that a slightly modified version of the CF strategy \cite{Cover1979}
achieves the optimal DM-tradeoff. More importantly, we show that
$d_{12}(r_{12},r_{21})$ $\left(d_{21}(r_{12},r_{21})\right)$ does not depend on $r_{21}$
($r_{12}$) and the two-way relay
channel can be decoupled into two one-way relay channels using the CF strategy.
Then we consider the more interesting case of half-duplex nodes, where the
achievable rate regions are protocol dependent. For the two-way relay channel
it is not known which protocol achieves the highest possible rates
\cite{Kim2007,Kim2008,Boche2007}. We use a three phase protocol, where in phase one $T_1$ transmits
to both the relay and $T_2$, in phase two $T_2$ transmits
to both the relay and $T_1$ and in phase three the relay transmits to $T_1$
and $T_2$. This three phase protocol makes use all the direct links between different nodes in a two-way relay channel. For this three phase protocol, we propose a
modified CF strategy and show that it can achieve the optimal DM-tradeoff in
some cases. We conjecture that our strategy can also
achieve the optimal DM-tradeoff in general, but we are yet to prove it.

{\it Notation:} The following notation is used in this paper. The
superscripts $^T, ^*$ represent the transpose and transpose
conjugate. ${\bf M}$ denotes a matrix, ${\bf m}$ a vector and $m_i$
the $i^{th}$ element of ${\bf m}$. For a matrix ${\bf M} = [{\bf
m}_1 \ {\bf m}_2 \ \ldots \ {\bf m}_n]$ by ${\text vec}({\bf M})$ we
mean $[{\bf m}^T_1 \ {\bf m}^T_2 \  \ldots \ {\bf m}^T_n]^T$.
$det({\bf A})$ and $tr({\bf A})$ denotes the determinant and trace
of matrix ${\bf A}$, respectively. $\bbE$ denotes the expectation.
$|| \cdot ||$ denotes the usual Euclidean norm of a vector and
$|\cdot|$ denotes the absolute value of a scalar. ${\bf I}_m$ is a
$m\times m$ identity matrix. $|{\cal X}|$ is the cardinality of set
${\cal X}$. We use the usual notation for $u(x) = {\cal O}(v(x))$
if $|\frac{u(x)}{v(x)}|$ remains bounded, as $x \rightarrow \infty$.
$x \sim {\cal CN}(0,\sigma)$ means $x$ is a circularly symmetric
complex Gaussian random variable with zero mean and variance
$\sigma$ and $x|y \sim {\cal CN}(0,\sigma)$ means given $y$, $x$ is
a circularly symmetric complex Gaussian random variable with zero
mean and variance $\sigma$. ${\bbC}^{MN}$ denotes the set of
$M\times N$ matrices with complex entries. $x_n \xrightarrow{w.p. 1}
y$ denotes that the sequence of random variables $x_n$ converge to a
random variable $y$ with probability $1$. We use $a \mapequal{w.p.1}
b$ to denote equality with probability $1$ i.e. $Prob.(a=b) =1$  and
 $\mapright{w.p.1}$ is defined similarly.
$I(x;y)$ denotes the mutual information between $x$ and $y$ and $h(x)$ the differential entropy of $x$ \cite{Cover2004}. To define a variable we use the symbol
$\bydef$.

{\it Organization:}
The rest of the paper is organized as follows. In Section \ref{sec:sys},
we describe the  two-way relay channel system model, the protocol under consideration and the key assumptions. In Section \ref{sec:optaf}, we obtain the optimal AF strategy to maximize the achievable rate region of the two-way relay channel. In Section \ref{sec:dcm}, we introduce a simple AF strategy, dual channel matching, and lower bound the achievable rate region of the optimal AF strategy of Section \ref{sec:optaf}. In Section \ref{sec:upbound}, we derive an upper bound on the capacity of the two-way relay channel capacity and compare it with the achievable rate region of the optimal AF strategy and dual channel matching. In Section \ref{sec:dmt}, we show that the CF strategy can achieve the optimal DM-tradeoff for full-duplex two-way relay channel, in general, and in some cases for the half-duplex case.
Final conclusions are made in Section \ref{conc}.

\section{System and Channel Model}
\label{sec:sys}
In this section we describe the two-way relay channel system model under consideration, and then present the relevant signal and channel models.
\begin{figure}
\centering
\includegraphics[height= 3in]{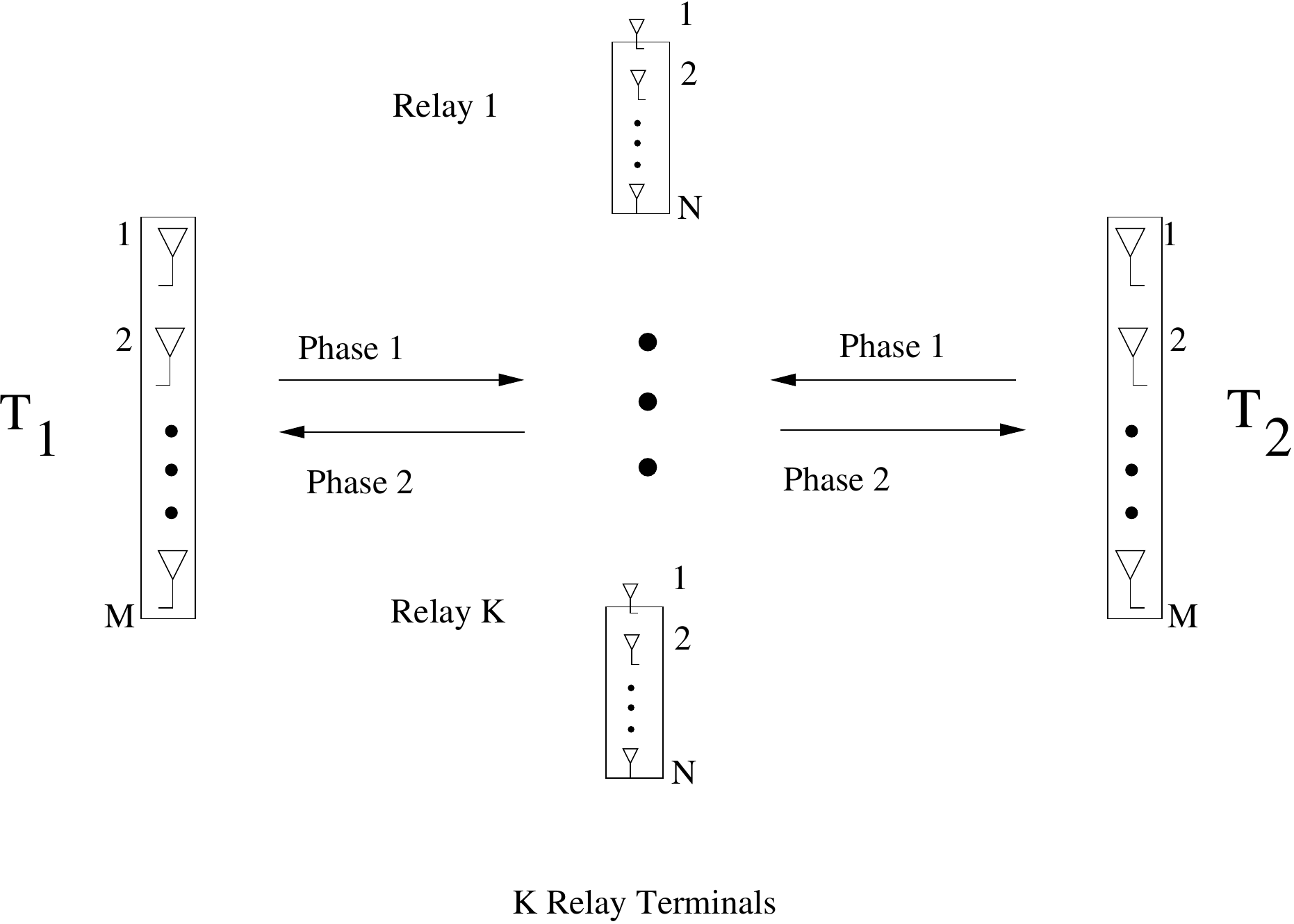}
\caption{Two-way relay channel system model with two phase communication}
\label{twoway}
\end{figure}
\subsection{System Model}
For the first part of the paper Section \ref{sec:optaf}, \ref{sec:dcm},
and \ref{sec:upbound}, we
consider a wireless network where there are two terminals $T_1$ and
$T_2$ who want to exchange information via $K$ relays, as shown
in Fig. \ref{twoway}. The $K$ relays do not have any data of their
own and only help $T_1$ and $T_2$ communicate. The $K$ relays are assumed to be
located randomly and independently so that the channel coefficients  between each relay and $T_1$ and $T_2$ are independent.
We also assume that there is no direct path between $T_1$ and $T_2$ and
that they can communicate only through the $K$ relays. This is a
realistic assumption when relaying is used for coverage improvement
in cellular systems, since at the cell edge the signal to noise ratio is extremely low for the direct path. In ad-hoc networks, it can be the case that two terminals want to communicate, but are out of each other's
transmission range.

We assume that both the terminals $T_1$ and $T_2$ have $M$ antennas
and all the $K$ relays have $N$ antennas each.
We further assume that both the terminals and all the relays can operate
only in half-duplex mode (cannot transmit and receive at the same time).
The communication protocol is summarized as follows  \cite{Rankov2005}.
In any given time slot, for the first $\alpha$ fraction of time, called the
{\it transmit phase}, both $T_1$ and $T_2$ are scheduled to transmit and
all the relays receive a superposition of the signals transmitted from $T_1$
and $T_2$. In the rest $(1-\alpha)$ fraction of the time slot, called the {\it receive phase},
all the relays are scheduled to
transmit simultaneously and both the terminals receive. Both $T_1$ and
$T_2$ are assumed to have power constraint of $P$, while for relays
we assume two different power constraints, the sum power constraint where the
sum of the power of all relays is $\le P_R$ or the individual power
constraint where each relay has power constraint of $P_R$.

\begin{figure}
\centering
\includegraphics[height= 2in]{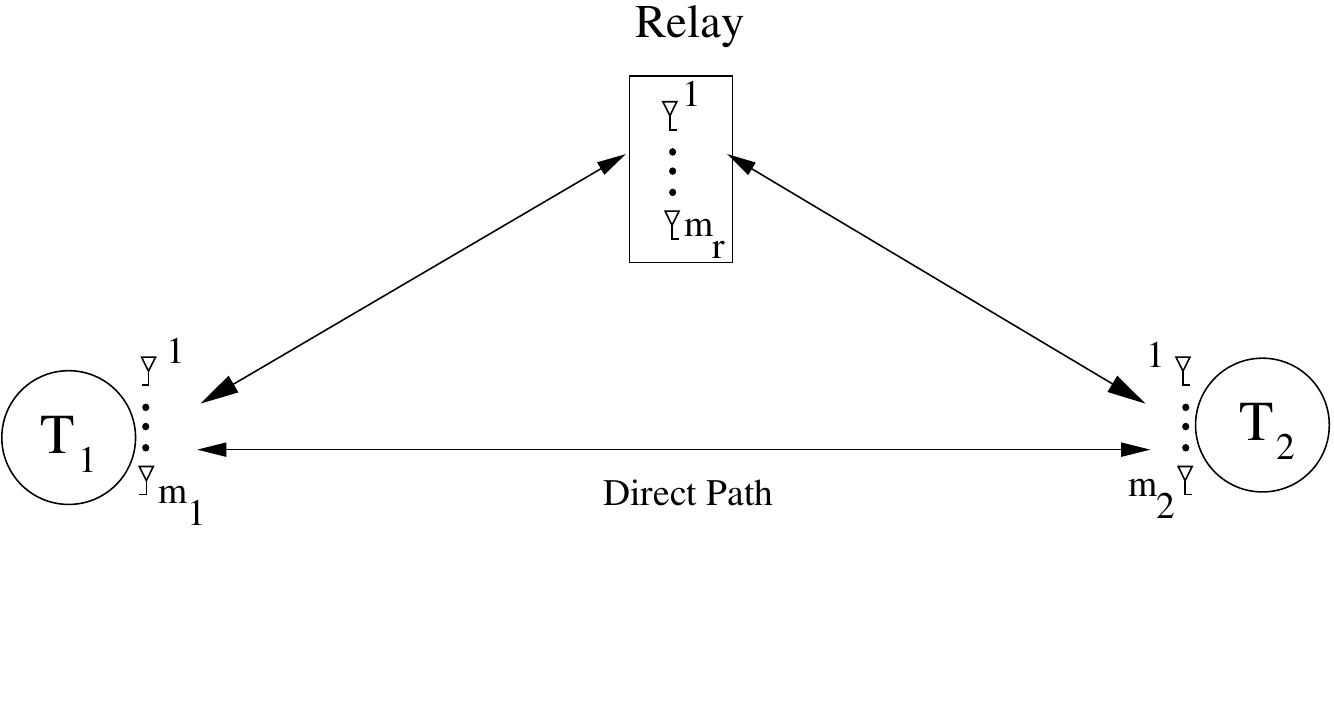}
\caption{System Model of the two-way relay channel with direct path for Section \ref{sec:dmt}}
\label{blkdmt}
\end{figure}

For the second part of the paper, Section \ref{sec:dmt}, we assume a two-way
relay channel with a single relay node and the presence of a direct path
between $T_1$ and $T_2$ as shown in Fig. \ref{blkdmt}. We assume that $T_1$ has $m_1$ antennas,
$T_2$ has $m_2$ antennas, and the relay node has $m_r$ antennas.

\begin{figure}
\centering
\includegraphics[height= 1.5in]{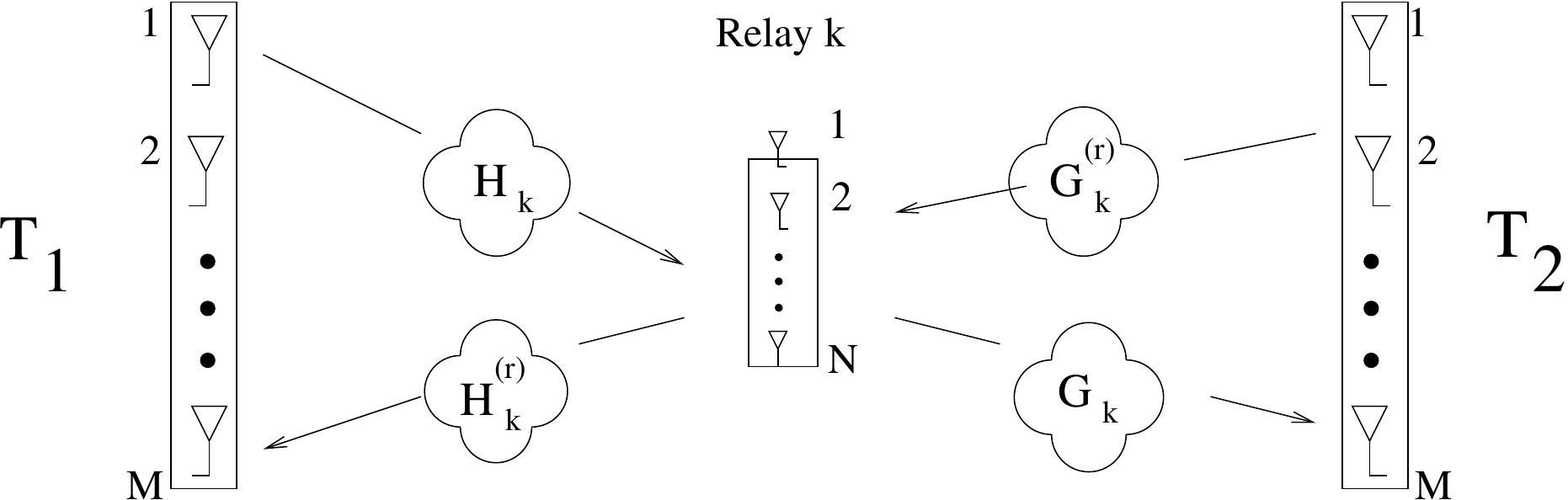}
\caption{Channel model for the two-way relay channel between $T_1$, $T_2$ and relay $k$}
\label{channel}
\end{figure}

\subsection{Channel and Signal Model}
Throughout this paper we assume that all the channels are
frequency flat slow fading block fading channels, where in a block of
time duration $T_c$ (called the coherence time), the channel coefficients
remain constant and change independently from block to block.
We assume that $T_c$ is more that the duration of time slot used by $T_1$ and $T_2$ to communicate with each other as described before.
As shown in Fig. \ref{channel}, let
the forward channel between $T_1$ and the $k^{th}$
relay be ${\bf H}_k = [{\bf h}_{1k} \ {\bf h}_{2k} \ \ldots \ {\bf h}_{Mk}]$
and the backward channel between $k^{th}$ relay and $T_1$
be ${\bf H}_k^{r} = [{\bf h}^{r}_{k1} \ {\bf h}^{r}_{k2} \ \ldots \ {\bf h}^{r}_{kM}]$.
Similarly let the  forward channel between $k^{th}$
relay and $T_2$ be ${\bf G}_k = [{\bf g}_{k1} \ {\bf g}_{k2} \ \ldots \ {\bf g}_{kM}]$
and the backward channel between $T_2$ and the $k^{th}$
relay be ${\bf G}_k^{r} =  [{\bf g}^{r}_{1k} \ {\bf g}^{r}_{2k}
 \ \ldots \ {\bf g}^{r}_{Mk}]$. For Section \ref{sec:dmt}, where the direct
path between $T_1$ and $T_2$ is considered, the channel between
$T_1$ and $T_2$ is denoted by $\bH_{12}$ and in the reverse direction by
$\bH_{12}^{r}$.
We assume that ${\bf H}_k, {\bf G}_k^{r} \in {\bbC}^{N\times
M}, {\bf H}_k^{r}, {\bf G}_k \in {\mathbb C}^{M\times N}, \bH_{12} \in
\bbC^{m_2\times m_1},  \bH_{12}^{r} \in \bbC^{m_1\times m_2}$ with independent and
 identically
distributed (i.i.d.) ${\cal CN}(0,1)$ entries.

For the first part of the paper Section \ref{sec:optaf}, \ref{sec:dcm} and \ref{sec:upbound},
we consider the following signal model.
The $N\times 1$ received signal at the $k^{th}$ relay is given by
\begin{equation}
\label{relayrx}
{\bf r}_k =  \sqrt{\frac{P}{M}} {\bf H}_{k}{\bx_1} + \sqrt{\frac{P}{M}}{\bf G}_{k}^{r}{\bx_2} + {\bf n}_k
\end{equation}
if ${\bf x}_1$ and ${\bx_2}$ are the $M\times 1$ signals transmitted from $T_1$ and $T_2$  to
be decoded at $T_2$ and $T_1$ respectively, with ${\bbE }\{{\bx_1}^{*}{\bx_1}\} = {\bbE }\{{\bx_2}^{*}{\bx_2}\} = M$, $P$ is the power transmitted by $T_1$ and  $T_2$, respectively. The noise $ {\bf n}_k$ is the $N\times 1$
spatio-temporal white complex Gaussian noise independent across
relays with ${\bbE}({\bf n}_k{\bf n}_k^*) = \ {\bf I}_N$.
Relay $k$ processes its incoming signal to transmit a $N\times 1$ signal
${\bf t}_k = \bW_k\br_k$ with $\sum_{k=1}^K{\bbE }\{{\bf t}_k^{*}{\bf t}_k\} \le P_R $ (sum power constraint) or ${\bbE }\{{\bf t}_k^{*}{\bf t}_k\} \le P_R $
(individual power constraint) in the receive phase.
The $M \times 1$ received signals
$\by_1$ and $\by_2$ at terminal $T_1$ and
$T_2$, respectively, in the receive phase, are given by
\begin{equation}
\label{t1rx}
\by_1 = \sum_{k=1}^{K}{\bf H}_k^{r}{\bf t}_k + {\bf z}_1,
\end{equation}
\begin{equation}
\label{t2rx}
\by_2 = \sum_{k=1}^{K}{\bf G}_k{\bf t}_k + {\bf z}_2,
\end{equation}
where ${\bf z}_1$ and ${\bf z}_2$
are $M\times 1$ spatio-temporal white complex Gaussian noise vectors with
${\bbE}({\bf z}_1{\bf z}_1^*) = {\bbE}({\bf z}_2{\bf z}_2^*) = \ {\bf I}_M$.

Throughout this paper we assume that both $T_1$ and $T_2$ perfectly know
$\{{\bf H}_k, {\bf H}_k^{r}, {\bf G}_k, {\bf G}_k^{r}\} \ \forall  \ k,$ $ k=1,2,,\ldots K$
in the receive mode. To be precise, in the receive phase (i.e. when $T_1$ and $T_2$ receive
signal from all the relays),
$T_1$ and $T_2$ both know $\{{\bf H}_k, {\bf G}_k\}$
and $\{{\bf H}_k^{r}, {\bf G}_k^{r}\}$  $ \forall  \ k, \ k=1,2,,\ldots K$.
We also assume that no transmit CSI is available at $T_1$ and $T_2$,
i.e. in the transmit phase $T_1$ and $T_2$ have
no information about what the realization of ${\bf H}_k$ and $ {\bf G}_k$ is going to be when it transmits its signal to all the relays in the transmit phase, respectively.

In this paper we assume different CSI assumptions at the
relay. For finding the optimal AF strategy (Section \ref{sec:optaf}) we assume that
each relay knows ${\bf H}_k, {\bf G}_k^{r}, {\bf G}_k, {\bf H}_k^{r}$ for
all $k=1,2\ldots,K$. To reduce the CSI requirements next, we present a
simple AF strategy in Section \ref{sec:dcm} where we assume that relay $k$ only knows
${\bf H}_k, {\bf G}_k^{r}, {\bf G}_k, {\bf H}_k^{r}$.
In Section \ref{sec:dmt}, we assume that the relay knows
${\bf H}_1, {\bf G}_1^{r}, {\bf G}_1, {\bf H}_1^{r}$, as well as
$\bH_{12}$, the
channel coefficient between $T_1$ and $T_2$.

\section{Optimal AF strategy for two-way relay channel}
\label{sec:optaf}
In this section we will find optimal relay beamformers that
maximize the achievable rate region of the two-way relay channel with AF, when $T_1$ and $T_2$ have a single antenna each, $M=1$.
For simplicity of exposition, in this section we consider the case when each relay nodes has a single antenna, $N=1$. Generalizations to $N >1$ are straightforward, and will be described later.

To start with, because of single antenna restriction, the channel between
$T_1$ and relay $k$ is denoted by $h_k$ and between relay $k$ and $T_2$ denoted by $g_k$.
For the reverse direction the channel coefficients are the same as in forward
direction but with an added superscript $r$, e.g. channel coefficient between
relay $k$ and $T_1$ is denoted by $h_k^r$. With AF strategy, each relay node transmits
the received signal multiplied with $w_k$ to both $T_1$ and $T_2$.
Thus, if $x_1$ and $x_2$ is the transmitted signal from $T_1$ and $T_2$,
respectively, then the received signal at $T_1$, $y_1$, and $T_2$, $y_2$ is
\begin{eqnarray}
\nonumber
y_1 &=& \sum_{k=1}^K \sqrt{P}h_k^rw_kg_k^rx_2 + \sqrt{P}h_k^rw_kh_kx_1 + h_k^rw_kn_k + z_1, \\ \label{rxsigscalar}
y_2 &=& \sum_{k=1}^K \sqrt{P}g_kw_kh_kx_1 + \sqrt{P}g_kw_kg_k^rx_1 + g_kw_kn_k + z_2,
\end{eqnarray}
where $n_k, \ \forall \ k=1,\ldots,K$ is ${\cal CN}(0,1)$ noise added at
relay $k$ and $z_1$, and $z_2$ are ${\cal CN}(0,1)$ added at $T_1$ and $T_2$.
Since $x_1$ and $x_2$ are known at $T_1$ and $T_2$, respectively, their
contribution can be removed from the received signal at $T_1$ and $T_2$,
respectively.
Let the rate of transmission from $T_1$ to $T_2$ be $R_{12}$ and
from $T_2$ to $T_1$ be $R_{21}$, then from (\ref{rxsigscalar})
\begin{eqnarray*}
R_{12} &=& \log \left(1+\frac{P\left(\left|\sum_{k=1}^Kg_kw_kh_k\right|^2\right)}
{1+\sum_{k=1}^K|g_kw_k|^2}\right), \\
R_{21} &=& \log \left(1+\frac{P\left(\left|\sum_{k=1}^Kh_k^rw_kg_k^r\right|^2\right)}
{1+\sum_{k=1}^K|h_k^rw_k|^2}\right).
\end{eqnarray*}
Thus, the achievable rate region for the two-way relay channel with AF for a
sum power constraint across all relays,
i.e. $p_R= P\sum_{k=1}^K(|w_kh_k|^2 + |w_kg_k^r|^2) +\sum_{k=1}^K|w_k|^2 \le P_R$ is the set
${\cal R}(P,P_R) = \cup_{p_R\le P_R}(R_{12}, R_{21})$
and for individual power constraint at each relay, i.e.
$p_{kR}= P(|w_kh_k|^2 + |w_kg_k^r|^2) +|w_k|^2 \le P_R$ is the set
${\cal R}(P,P_R) = \cup_{p_{kR}\le P_R, \ k=1,\ldots,K}(R_{12}, R_{21})$.
Therefore, the problem is to find optimal $w_k$'s
that achieve the boundary points of the region ${\cal R}(P,P_R)$, for both
the sum power constraint and an individual power constraint.

For the one-way relay channel, no communication from $T_2$ to $T_1$,
optimal $w_k$'s have been found in \cite{Yi2007} to maximize $R_{12}$.
The solution of \cite{Yi2007}, provides an upper bound on individual
rates $R_{12}$ and $R_{21}$ and is equivalent to solutions where
$R_{12}$ or $R_{21}$ is greedily maximized disregarding the other.
The problem in the two-way relay channel case is to find optimal $w_k$'s
such that $R_{sum} = R_{12} + R_{21}$ is
maximized, for each $\beta \in [0,1]$, where $R_{12} = \beta R_{sum}$, and
$R_{21} = (1-\beta)R_{sum}$.
Towards that end, we use the rate profile method \cite{Cioffi2006}
to identify $w_k$'s that meet the boundary point of ${\cal R}(P,P_R)$.
Next, we only consider the sum power constraint across the relays.
For individual power constraints the same procedure can be applied as pointed
out later.
Thus, the optimization problem can be formulated as follows.
\begin{equation}
\begin{tabular}{cl}
\label{origafopt}
Maximize$_{w_k, \ k=1,2,\ldots,K}$ & $R_{sum}$ \\
{subject to}   & $\log \left(1+\frac{P\left(\left|\sum_{k=1}^Kg_kw_kh_k\right|^2\right)} {1+\sum_{k=1}^K|g_kw_k|^2}\right) \ge \beta R_{sum},$  \\
 & $\log \left(1+\frac{P\left(\left|\sum_{k=1}^Kh_k^rw_kg_k^r\right|^2\right)}
{1+\sum_{k=1}^K|h_k^rw_k|^2}\right) \ge (1-\beta) R_{sum}$, \\
& $P\sum_{k=1}^K(|w_kh_k|^2 + |w_kg_k^r|^2) + \sum_{k=1}^K|w_k|^2 \le P_R.$
\end{tabular}
\end{equation}

An equivalent problem to this problem is the following iterative
power minimization problem subject to rate constraints,
\begin{equation}
\begin{tabular}{cl}
\label{equiiterafopt}
      Minimize$_{w_k, \ k=1,2\ldots,K}$ &  $p_R = P\sum_{k=1}^K(|w_kh_k|^2 + |w_kg_k^r|^2) + \sum_{k=1}^K|w_k|^2$ \\
      subject to   & $\log \left(1+\frac{P\left(\left|\sum_{k=1}^Kg_kw_kh_k\right|^2\right)}
{1+\sum_{k=1}^K|g_kw_k|^2}\right) \ge \beta R_{sum}^u$,  \\
      & $\log \left(1+\frac{P\left(\left|\sum_{k=1}^Kh_k^rw_kg_k^r\right|^2\right)}{1+\sum_{k=1}^K|h_k^rw_k|^2}\right) \ge (1-\beta)R_{sum}^u$,
    \end{tabular}
\end{equation}
where at each iteration $R_{sum}^u$ is changed to maximize the achievable
rate, subject to power constraint. To be precise, if the value of $R_{sum}^u$
at iteration $i$ is say $x$ and the solution to (\ref{equiiterafopt}) is feasible
(i.e. if $p_R \le P_R$) \footnote{For an
individual power constraint the same can
be done by checking at each iteration whether the obtained solution $p_R$ is
feasible with individual power constraints or not.}, then $x$ is incremented in next iteration, otherwise
decreased. Choice of the step size of increase or decrease determines the
speed of convergence to the optimal rate $R_{sum}^u$, for which $p_R \le P_R$.
One possible starting point for $R_{sum}^u$ is $2$ times the maximum $R_{12}$
provided by \cite{Yi2007} for one way relay channel. The step size can be chosen by
bisection between the last feasible $R_{sum}^u$ (initially $0$) and the
last infeasible $R_{sum}^u$.
Even though this equivalent problem provides a solution to (\ref{origafopt}) in a
iterative manner, the problem (\ref{equiiterafopt}) is in general non-convex, and not easy to solve.
To overcome this limitation, we recast
the problem (\ref{equiiterafopt}) as a standard power minimization problem subject to
signal-to-interference-noise ratio (SINR) \cite{Wiesel2006}, where the forwarded noise
from each relay plays the role of interference. For a given $\beta$ and $R_{sum}^u$,
the problem (\ref{equiiterafopt}) is of the form
\begin{equation}
\label{powerminafopt}
\begin{tabular}{cl}
      Minimize$_{w_k, \ k=1,2\ldots,K}$ &  $p_R = P\sum_{k=1}^K(|w_kh_k|^2 + |w_kg_k^r|^2) + \sum_{k=1}^K|w_k|^2$ \\
      subject to   & $\frac{\left|\sum_{k=1}^Kg_kw_kh_k\right|^2}
{1+\sum_{k=1}^K|g_kw_k|^2} \ge \frac{2^{\beta R_{sum}^u}-1}{P} \bydef \gamma_0  $\\
      &  $\frac{\left|\sum_{k=1}^Kh_k^rw_kg_k^r\right|^2}
{1+\sum_{k=1}^K|h_k^rw_k|^2} \ge \frac{2^{(1-\beta)R_{sum}^u}-1}{P} \bydef \gamma_1$.
    \end{tabular}
\end{equation}
This problem again is non-convex, however, it is of the form
\begin{equation}
\label{weiselafopt}
    \begin{tabular}{cl}
      Minimize &  $f(x)$\\
      subject to   & $||a_i(x)||^2 - |b_i(x)|^2 \le 0, \ \forall \ i$,
    \end{tabular}
\end{equation}
% % \end{center}
%\end{displaymath}
where $f(x)$ is a convex function, $a_i(x)$ is an affine function of $x$ and $b_i(x) \ge 0$ $\ \forall \ i
$, by noting that if $\sum_{k=1}^Kg_kw_kh_k$, or $\sum_{k=1}^Kh_k^rw_kg_k^r$ are
less than zero or complex, then they can be scaled by appropriate phases to
make them real and positive, without
changing the objective function or the constraints
\footnote{An immediate consequence of this property is that the optimal
solution does not change if all $g_k's$ are scaled by $e^{j\phi_1}$, or
all $h_k^r$'s are scaled by $e^{j\phi_2}$.}.

For the problem (\ref{weiselafopt}), it has been shown in \cite{Wiesel2006}, that if the problem is strictly feasible, then KKT conditions \cite{Boyd2004} are necessary and sufficient to find the
optimal solution. It is easy to see that the problem (\ref{powerminafopt}) is strictly feasible and therefore KKT conditions are sufficient for optimality.
The Lagrangian of problem (\ref{powerminafopt}) is of the form
\begin{eqnarray*}
{\cal L} &=& \bw\bA\bw^* + \lambda_1\left(\bw\bB\bw^* - \frac{1}{\gamma_0} |\bc\bw^T|^2 + 1\right) + \lambda_2\left(\bw\bD\bw^* - \frac{1}{\gamma_1} |\bee\bw^T|^2 + 1\right),
\end{eqnarray*}
where $\bw = [w_1 \ \ldots \ w_K]$ and
\[\bA =\left[\begin{array}{cccc}P(|h_1|^2+|g_1^r|^2) + 1 & 0 & \ldots & 0 \\
0 & P(|h_2|^2+|g_2^r|^2) + 1 &  \ldots & 0 \\
0 & 0 & \ddots & 0 \\
0 & \ldots & 0 & P(|h_K|^2+|g_K^r|^2) + 1
\end{array}\right], \]
\[\bB =\left[\begin{array}{cccc}|g_1|^2 & 0 & \ldots & 0 \\
0 & |g_2|^2 &  \ldots & 0 \\
0 & 0 & \ddots & 0 \\
0 & \ldots & 0 & |g_K|^2
\end{array}\right], \
\bD =\left[\begin{array}{cccc}|h_1^r|^2 & 0 & \ldots & 0 \\
0 & |h_2^r|^2 &  \ldots & 0 \\
0 & 0 & \ddots & 0 \\
0 & \ldots & 0 & |h_K^r|^2
\end{array}\right], \] and
$\bc =\left[\begin{array}{ccc}g_1h_1 & \ldots  & g_Kh_K
\end{array}\right], \ \bee =\left[\begin{array}{ccc}h_1^rg_1^r & \ldots  & h_K^rg_K^r
\end{array}\right].$

Differentiating the Lagrangian yields
\[ \left(\bA + \lambda_1\bB+\lambda_2\bD -\frac{\lambda_1}{\gamma_0}\bc^{*}\bc+
\frac{\lambda_2}{\gamma_1} \bee^{*}\bee\right)\bw  =0, \] and the optimal $\bw$ is found by solving for
$\lambda_1$ and $\lambda_2$ using the constraints \footnote{Clearly, the
optimal $\bw$ lies in the null space of some matrix that is a function of
$\bA,\bB,\bc,\bee$, and $\bD$ and hence not unique.}.

Therefore, by recasting our original problem of obtaining the boundary
points of ${\cal R}(P,P_R)$ to the power minimization problem with
SINR constraints, we have shown that the optimal solution can be
found in an efficient way. In Section \ref{sec:upbound}, we plot the achievable
rate region of the optimal AF strategy and compare it with the lower bound
obtained by using dual channel matching, and an upper bound.

 Recall that we only considered a
two-way relay channel, where each relay had a single antenna, $N=1$. Extension to $N>1$, is straightforward by
replacing $g_kw_kh_k$ by $\bg_k\bW_k\bh_k$, $g_kw_k$ by $\bg_k\bW_k$,
$h_k^rw_kg_k$ by $\bh_k^r\bW_k\bg_k$ and $h_k^rw_k$ by $\bh_k^r\bW_k$, which are scalars as before, and the optimal solution to $\bW_k$'s can be found using the iterative power minimization algorithm (\ref{equiiterafopt}).

Our algorithm to optimize the achievable region with AF
is fairly simple, however, it assumes that each relay has CSI for all the relay nodes, and requires $M=1$.
Finding optimal relay beamformers where each relay has only its CSI, and $M>1$, is rather
complicated and has not been solved even for the one-way relay channel \cite{Yi2007}.
Another limitation of the optimal AF strategy is that the expression for the
obtained rate region cannot be written down in close form, and therefore does not allow
analytical tractability for comparison with an upper bound.
To remove  these restrictions, in the next section we propose a simple
AF strategy, called dual channel matching,
where each relay uses its own CSI, and for which the achievable rate region
expression can be written down in a closed form. Since dual channel matching
is in general, a suboptimal AF strategy, the achievable rate region of dual
channel matching lower bounds the rate region of the optimal AF strategy, and
allows to estimate the difference between the optimal AF strategy and the upper bound.

\section{Dual Channel Matching Strategy}
\label{sec:dcm} In this section we propose a simple AF strategy,
called dual channel matching, and derive a lower bound on the achievable
rate region for the two-way relay channel. With the dual
channel matching strategy relay $k$ multiplies
$\sqrt{\beta_k} \left(\bG_k^*\bH_k^* + {\bf H}_k^{r*}{\bf
G}^{r*}_{k}\right)$ to the received signal and forwards it to $T_1$
and $T_2$, where $\beta_k$ is the normalization constant to satisfy
the power constraint. Dual channel matching tries to match both the
channels which the data streams from $T_1$ to $T_2$ and $T_2$ to
$T_1$ experience at each relay node. The motivation for this
strategy is that for one-way relay channel (i.e. $T_2$ has no data
for $T_1$) with one relay node, the optimal AF strategy is to
multiply $\bV_2\bD\bU_1^{*}$ to the signal at the relay, where the
singular value decomposition of ${\bf H}_1$ is $ \bU_1\bD_1\bV_1^*$
and ${\bf G}_1$ is  $\bU_2\bD_2\bV_2^*$ and $\bD$ is a diagonal
matrix whose entries are chosen by waterfilling
\cite{Munoz-Medina2007}. In dual channel matching the complex
conjugates of the channels are used directly rather than the unitary
matrices from the SVD of the channels \cite{Munoz-Medina2007}. This
modification makes it easier to analyze the achievable rates for the
two-way relay channel. Note that the dual channel matching is an
extension of the listen and transmit strategy of \cite{Dana2006} for
the one-way relay channel, where each relay transmits the received
signal after scaling it with the complex conjugates of the forward
and backward channel coefficients.

Together with dual channel matching we
restrict the signal transmitted from $T_1$ and $T_2$, $\bx_1$ and
$\bx_2$, respectively, to be circularly symmetric complex Gaussian distributed with
covariance matrix ${\bbE}\{{\bx_1}{\bx_1}^*\} =
{\bbE}\{{\bx_2}{\bx_2}^*\} = \bI_{M}$, to obtain a lower bound
on the achievable rate region of  two-way relay channel.
Moreover, we use $\alpha = \frac{1}{2}$ i.e. $T_1$ and $T_2$ transmit and
receive for same amount of time. The achievable rates $R_{12}$ and $R_{21}$ using the dual channel matching can be computed as follows.

From (\ref{relayrx}), the received signal at the $k^{th}$ relay is given by
\[{\bf r}_k =  \sqrt{\frac{P}{M}} {\bf H}_{k}{\bx_1} +
\sqrt{\frac{P}{M}}{\bf G}^{r}_{k}{\bx_2} + {\bf n}_k.\]

Using dual channel matching as described above, at relay $k$,
$\bG_k^*\bH_k^* + {\bf H}_k^{r*}{\bf G}^{r*}_{k}$ is multiplied to the
received signal so that the transmitted signal ${\bf t}_k$ is
given by
\[{\bf t}_k =\sqrt{\beta_k}
\left(\bG_k^*\bH_k^* + {\bf H}_k^{r*}{\bf G}^{r*}_{k}\right)\br_k\]
where $\beta_k$ is to ensure that $\sum_{k=1}^K{\bf t}_k^*{\bf t}_k
=P_R$ \footnote{This is for the sum power constraint. For an
individual power constraint, $\beta$ is chosen such that ${\bf
t}_k^*{\bf t}_k =P_R$ for each $k$.}. With dual channel matching the
received signal at $T_2$ is given by
\begin{equation}
\label{destrecsig}
\by_2 = \sum_{k=1}^K\bG_k\bt_k + \bz.
\end{equation}
Expanding (\ref{destrecsig}) we can write
\begin{eqnarray*}
\by &=& \underbrace{\sum_{k=1}^K\sqrt{\frac{P\beta_k}{M}}\bG_k
\left(\bG_k^*\bH_k^* + {\bf H}_k^{r*}{\bf G}^{r*}_{k}\right)\bH_k}_{\bA}\bx_1 +
\sum_{k=1}^K\sqrt{\frac{P\beta_k}{M}}\bG_k
\left(\bG_k^*\bH_k^* + {\bf H}_k^{r*}{\bf G}^{r*}_{k}\right){\bf G}^{r}_{k}\bx_2\\
&&+
\sum_{k=1}^K\underbrace{\sqrt{\beta_k}\bG_k
\left(\bG_k^*\bH_k^* + {\bf H}_k^{r*}{\bf G}^{r*}_{k}\right)}_{\bB_k}{\bf n}_k
+ \bz.
\end{eqnarray*}
Since $\bx_2$ and all the channel coefficients are known at
$T_2$, the second term can be removed from the received signal at $T_2$.
Moreover, as described before $\bx_1$ is circularly symmetric
complex Gaussian vector
with covariance matrix $\bQ=\bI_M$, thus the achievable rate for
$T_1$ to $T_2$ link is
\cite{Telatar1999}
\begin{equation}
\label{lowerbddcm1}
R_{12}= \frac{1}{2}I(\bx_1;\by_2) =
\frac{1}{2}\log\det\left(\bI_M +  \bA\bA^*\left(\sum_{k=1}^K\bB_k\bB_k^* + \bI_M\right)^{-1}\right),\end{equation}
since $\bbE\left\{\bn_k\bn_k^*\right\} = \bbE\left\{\bz\bz^*\right\}= \bI_M, \ \forall \ k.$
Similarly, we obtain the expression for $R_{21}$,
\begin{equation}
\label{lowerbddcm2}R_{21}= \frac{1}{2}\log\det\left(\bI_M +  \bC\bC^*
\left(\sum_{k=1}^K\bD_k\bD_k^* + \bI_M\right)^{-1}\right),
\end{equation}
where $\bC = \sum_{k=1}^K\sqrt{\frac{P\beta_k}{M}}\bH^{r}_k
\left(\bG_k^*\bH_k^* + {\bf H}_k^{r*}{\bf
G}^{r*}_{k}\right)\bG^{r}_k$ and $\bD_k =
\sqrt{\beta_k}\bH^{r}_k \left(\bG_k^*\bH_k^* + {\bf
H}_k^{r*}{\bf G}^{r*}_{k}\right).$ This rate region expression
obtained is analytically tractable and can be used to compare the
loss between the optimal AF strategy and the upper bound. Another
interesting question of interest is how does the achievable rate
region behaves with $K$. To answer that question, we turn to
asymptotics and compute the rate region in the limit $K\rightarrow
\infty$, in the next lemma.
\begin{lemma} As $K$ grows large, $K \rightarrow \infty$,
\begin{eqnarray*}
\lim_{K \rightarrow \infty}R_{12} &\mapequal{w.p.1} &
\frac{M}{2}\log K + {\cal O}(1),\\
\lim_{K \rightarrow \infty}R_{21} &\mapequal{w.p.1}& \frac{M}{2}\log K + {\cal O}(1).
\end{eqnarray*}
\end{lemma}
\begin{proof}
Consider
\begin{eqnarray*}
2R_{12}- \log\det K\bI_M &=& \log\det\left(\bI_M +  \bA\bA^*
\left(\sum_{k=1}^K\bB_k\bB_k^* + \bI_M\right)^{-1}\right) - \log\det K\bI_M, \\
&=& \log\det\left(\frac{1}{K}\bI_M +  \frac{\bA}{\sqrt{K}}\frac{\bA^*}{\sqrt{K}}\left(\sum_{k=1}^K\bB_k\bB_k^* + \bI_M\right)^{-1}\right).
\end{eqnarray*}
To satisfy the sum power constraint, let $\beta = \frac{P_R}{c_1K}$ \footnote{Equal power allocation among relays.}, where
$c_1$ is a constant such that
\[c_1 =
\bbE\left\{
\left(
     \left(\bG_k^*\bH_k^* + {\bf H}_k^{r*}{\bf G}^{r*}_{k}\right)\br_k\right)^*
\left(\bG_k^*\bH_k^* + {\bf H}_k^{r*}{\bf G}^{r*}_{k}\right)\br_k
\right\},\]
which is same for all $k$.
Then,
\[\frac{\bA}{\sqrt{K}} = \sqrt{\frac{PP_R}{c_1M}}\frac{1}{K}\sum_{k=1}^K\bG_k
\left(\bG_k^*\bH_k^* + {\bf H}_k^{r*}{\bf G}^{r*}_{k}\right)\bH_k, \]
which by using strong law of large numbers, converges to,
\[\frac{{\bA}}{\sqrt{K}} \xrightarrow{w.p.1} \sqrt{\frac{PP_R}{c_1M}} N^2\bI_M,\]
since $\bbE\left\{\bG_k\bG_k^*\right\} = \bbE\left\{\bH_k^*\bH_k\right\} = N\bI_M \ \forall \ k$, and
$\bbE\left\{\bG_k\bH_k^{r*}\right\}  = 0\bI_M\ \forall \ k$.
Same result holds true for $\frac{\bA^*}{\sqrt{K}}$.
With $\beta =\frac{P_R}{c_1K}$,
\[\sum_{k=1}^K\bB_k\bB_k^* = \frac{P_R}{c_1}
\frac{1}{K}\sum_{k=1}^K\left(\bG_k
\left(\bG_k^*\bH_k^* + {\bf H}_k^{r*}{\bf G}^{r*}_{k}\right)\right)\left(\bG_k
\left(\bG_k^*\bH_k^* + {\bf H}_k^{r*}{\bf G}^{r*}_{k}\right)\right)^*,\]
which again using the strong law of large numbers converges to
$\frac{P_R}{c_1}\theta\bI_M$,
for some finite $\theta$,
since $\bH_k, \bG_k, \bH_k^r, \bG_k^r{\bB}_k^*$ are i.i.d. with finite variance.
Thus, in the limit $K \rightarrow \infty$,
\[2R_{12}- \log\det K\bI_{M} \rightarrow M\log\left(\frac{PP_RN^4c_1}{M(P_R\theta+c_1)}\right), \] and thus
it follows that
\begin{equation}
\label{lbdcmass1}R_{12} \mapequal{w.p.1}  \frac{M}{2}\log K + {\cal O}(1).
\end{equation}
Similarly we get the achievable rate $R_{21}$ on the $T_2$ to $T_1$ link
as
\begin{equation}
\label{lbdcmass2}
\lim_{K \rightarrow \infty}R_{21} \mapequal{w.p.1} \frac{M}{2}\log K + {\cal O}(1).\end{equation}
\end{proof}

\vspace{0.25in} {\it Discussion:}
In this section we introduced the
dual channel matching AF strategy, and obtained a lower bound on the
capacity region of the two-way relay channel.
Dual channel matching is a simple AF strategy that requires local CSI, and as
we will see in Section \ref{sec:upbound}, has achievable rate region
very close to that of the optimal AF strategy
(Section \ref{sec:optaf}) for $M=1$.
We also derived the asymptotic achievable rate region of the dual channel matching, by taking the limit $K \rightarrow \infty$, and using the law of large numbers. We showed, that in the asymptotic regime,
both $R_{12}$ and $R_{21}$ scale as $\frac{M}{2}\log K$ with increasing $K$.

Next, we derive an upper bound on the capacity region of the two-way relay
channel, and compare it with the achievable rate region of the dual channel matching.

\section{Upper Bound on the Two-Way Relay Channel Capacity}
\label{sec:upbound}
\begin{figure}
\centering
\includegraphics[height= 2in]{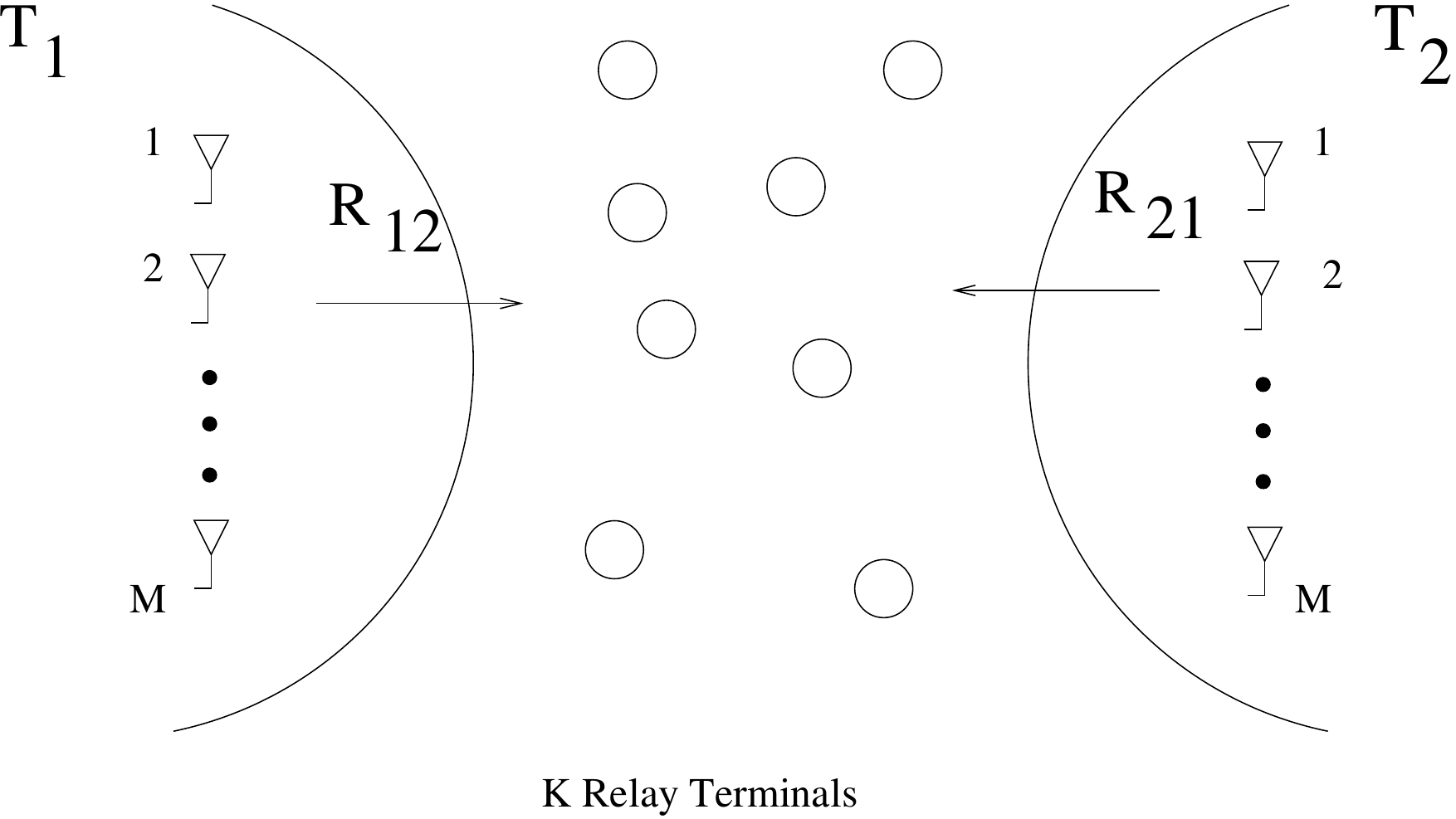}
\caption{Broadcast Cut}
\label{cutset}
\end{figure}
In this section we upper bound the capacity region of the two-way relay
channel using the cut-set bound \cite{Cover2004} for the broadcast cut, and the
multiple access cut. We assume a general two-phase protocol where for
$\alpha$ fraction of the time slot $T_1$ and $T_2$ transmit to all relays and
the rest of the $(1-\alpha)$ fraction of time slot all relays
simultaneously transmit to both $T_1$ and $T_2$. Note that to lower bound
the capacity of the two-way relay channel using dual channel matching, we
used $\alpha = \frac{1}{2}$ which might be suboptimal. We prove later
that for the asymptotic case of $K \rightarrow \infty$,
$\alpha =  \frac{1}{2}$ is optimal.

\begin{figure}
\centering
\includegraphics[height= 2in]{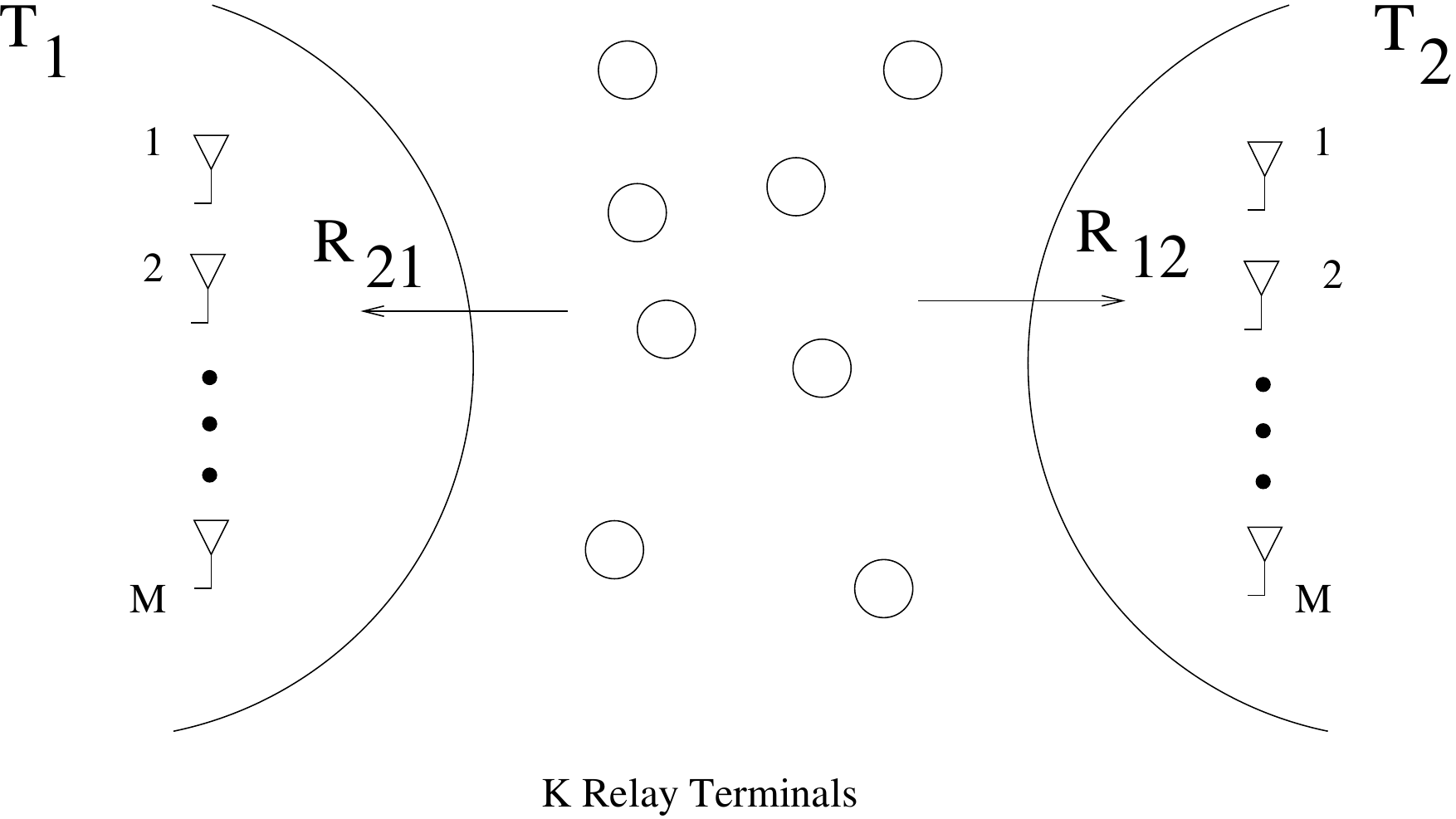}
\caption{Multiple Access Cut}
\label{cutsetmac}
\end{figure}

The upper bound is derived as follows.
We start by first separating $T_1$ and then $T_2$ from
the network and apply the cut set bound \cite{Cover2004} to upper bound the
rate of information
transfer between $T_1 \rightarrow T_2$ and $T_2 \rightarrow T_1$, respectively. Using the cutset bound, we first show that the maximum rate at
of information transfer  from
$T_1 \rightarrow T_2$ ($T_2 \rightarrow T_1$) is upper bounded
by the maximum rate of information transfer between $T_1$ ($T_2$)
and $r_1, r_2, \ldots, r_K$ (broadcast cut) and also by the
maximum rate of information transfer between
$r_1, r_2, \ldots, r_K$ and $T_2$ ($T_1$) (multiple access cut),
Fig. \ref{cutset} and \ref{cutsetmac}.
Then we use the capacity results
from \cite{Telatar1999} to upper bound the maximum rate through the broadcast cut
for the case when CSI is only available at the receiver (all relays)
and all the relays collaborate to decode the information.
Similarly, for the multiple access cut as shown in Fig. \ref{cutsetmac},
we upper bound the maximum rate at which
all the $r_1, r_2, \ldots r_K$ can communicate to $T_2$ ($T_1$) by
using capacity results from \cite{Telatar1999}, when CSI is known both at the
transmitter (all relays) and the receiver ($T_1, T_2$) and
all the relays collaborate to transmit the information.

{\bf Broadcast cut} - To derive an upper bound we make use of the cutset bound (Section 14.10 \cite{Cover2004}).
Separating the terminal $T_1$ from the rest of the network and applying
the cutset bound on the broadcast cut as shown in Fig. \ref{cutset},
\begin{equation}
\label{uppbound1}
R_{12} \le \alpha\left\{I({\bx_1};{\bf r}_1, {\bf r}_2,\ldots , {\bf r}_K,
\by_2 | {\bf t}_1, {\bf t}_2, \ldots {\bf t}_K, {\bx_2})\right\}.
\end{equation}
Again applying the cutset bound while separating the terminal $T_2$,
\begin{equation}
\label{uppbound2}
R_{21} \le \alpha\left\{I({\bx_2};{\bf r}_1,{\bf r}_2,\ldots ,{\bf r}_K, \by_1 | {\bf t}_1, {\bf t}_2, \ldots {\bf t}_K, {\bx_1})\right\}
\end{equation}
for some joint distribution $p({\bx_1},{\bf t}_1,{\bf t}_2,\ldots,{\bf t}_K, {\bx_2})$, where $R_{12}$ and $R_{21}$ are the maximum rates at which $T_1$ can
communicate to $T_2$ and $T_2$ can communicate to $T_1$ respectively, reliably.
By the definition of mutual information \cite{Cover2004}
%\begin{eqnarray*}
%I(A;B,C|D) & = & H(A|D) - H(A|B,C,D) \\
%&= &H(A|D)- H(A|C,D) + H(A|C,D) -H(A|B,C,D) \\
%&= &I(A;C|D) + I(A;B|C,D) \\
%\end{eqnarray*}
%for any $A,B,C,D$.
%and it follows that
\begin{eqnarray}\nonumber
I({\bx_1};{\bf r_1},{\bf r}_2,\ldots ,{\bf r}_K, \by_2 | {\bf t}_1, {\bf t}_2, \ldots {\bf t}_K, {\bx_2})  & = &
I({\bx_1};{\bf r}_1,{\bf r}_2,\ldots ,{\bf r}_K | {\bf t}_1, {\bf t}_2, \ldots {\bf t}_K, {\bx_2}) \\\label{itdummy1}
& & + \  I({\bx_1};\by_2 |{\bf r}_1,{\bf r}_2,\ldots ,{\bf r}_K, \ {\bf t}_1, {\bf t}_2, \ldots {\bf t}_K, {\bx_2}).
\end{eqnarray}
By expanding the mutual information in terms of entropy,
\begin{eqnarray*}
I({\bx_1};{\bf r}_1,{\bf r}_2,\ldots ,{\bf r}_K | {\bf t}_1, {\bf t}_2, \ldots {\bf t}_K, {\bx_2}) & = & h({\bx_1} | {\bf t}_1, {\bf t}_2, \ldots {\bf t}_K, {\bx_2}) \\
& & - \ h({\bx_1}| {\bf r}_1,{\bf r}_2,\ldots ,{\bf r}_K, {\bf t}_1, {\bf t}_2, \ldots {\bf t}_K, {\bx_2} )
\end{eqnarray*}
Since conditioning can only reduce entropy \cite{Cover2004},
\begin{eqnarray*}
I({\bx_1};{\bf r}_1,{\bf r}_2,\ldots ,{\bf r}_K | {\bf t}_1, {\bf t}_2, \ldots {\bf t}_K, {\bx_2}) & \le & h({\bx_1}|{\bx_2}) \\
& & - \ h({\bx_1}| {\bf r}_1,{\bf r}_2,\ldots ,{\bf r}_K, {\bf t}_1, {\bf t}_2, \ldots {\bf t}_K, {\bx_2} ).
\end{eqnarray*}
Note that ${\bf t}_1,{\bf t}_2,\ldots ,{\bf t}_K$ is a function of
${\bf r}_1,{\bf r}_2,\ldots ,{\bf r}_K$, which implies
\begin{eqnarray*}
I({\bx_1};{\bf r}_1,{\bf r}_2,\ldots ,{\bf r}_K | {\bf t}_1, {\bf t}_2, \ldots {\bf t}_K, {\bx_2}) & \le & h({\bx_1} |{\bx_2}) \\
& & - \ h({\bx_1}| {\bf r}_1,{\bf r}_2,\ldots ,{\bf r}_K, {\bx_2} )
\end{eqnarray*}
and hence\footnote{Without $\bx_2$, in \cite{Bolcskei2006}, this inequality has been shown to be an equality, which is incorrect.}
\begin{equation}
\label{itdummy2}
I({\bx_1};{\bf r}_1,{\bf r}_2,\ldots ,{\bf r}_K | {\bf t}_1, {\bf t}_2, \ldots {\bf t}_K, {\bx_2}) \le I({\bx_1};{\bf r}_1,{\bf r}_2,\ldots ,{\bf r}_K | {\bx_2}) .\end{equation}
Given perfect channel knowledge at terminal $T_2$,
\[I({\bx_1};\by_2 |{\bf r}_1,{\bf r}_2,\ldots ,{\bf r}_K, \  {\bf t}_1, {\bf t}_2, \ldots {\bf t}_K, {\bx_2}) = I({\bx_1},{\bf z_2})\]
where ${\bf z}_2$ is the AWGN noise. Since ${\bx_1}$ and ${\bf z}_2$
are independent, $I({\bx_1},{\bf z}_2) = 0$, and therefore from
(\ref{itdummy1}, \ref{itdummy2}),
\[I({\bx_1};{\bf r_1},{\bf r}_2,\ldots ,{\bf r}_K,\by_2 | {\bf t}_1, {\bf t}_2, \ldots {\bf t}_K, {\bx_2}) \le I({\bx_1};{\bf r}_1,{\bf r}_2,\ldots ,{\bf r}_K | {\bx_2}).\]
Hence from (\ref{uppbound1}),
\begin{equation}
\label{alphaupbound1}
R_{12} \le I({\bx_1};{\bf r}_1,{\bf r}_2,\ldots ,{\bf r}_K|{\bx_2}).
\end{equation}
Similarly,
by interchanging the roles of ${\bx_1}$ and ${\bx_2}$,
\begin{equation}
\label{alphaupbound2}
R_{21} \le
I({\bx_2};{\bf r}_1,{\bf r}_2,\ldots ,{\bf r}_K|{\bx_1}).
\end{equation}
Therefore it is clear that both $R_{12}$ and $R_{21}$
is upper bounded by the maximum
information flow through the broadcast cut Fig. \ref{cutset} when all the relays
are allowed to collaborate.
%Since we assume that the sources $ T_1$ and $T_2$ transmit only for
%$\alpha$ fraction of the time in each time slot,
%\begin{equation}
%\label{alphaupbound1}
%R_{12} \le {\bbE}_{\{{{\bf H}_k, {\bf G}_k\}}_{k=1}^K}\left\{\alpha I({\bx_1};{\bf r}_1,{\bf r}_2,\ldots ,{\bf r}_K|{\bx_2})\right\}.
%\end{equation}
%\begin{equation}
%\label{alphaupbound2}
%R_{21} \le {\bbE}_{\{{H^{r}_k,  {\bf G}^{r}_k\}}_{k=1}^K}\left
%\{\alpha
%I({\bx_2};{\bf r}_1,{\bf r}_2,\ldots ,{\bf r}_K|{\bx_1})\right\}.
%\end{equation}
Expanding the mutual information in terms of differential entropy,
\[ I({\bx_1};{\bf r}_1,{\bf r}_2,\ldots ,{\bf r}_K| {\bx_2}) =  h({\bf r}_1,{\bf r}_2,\ldots ,{\bf r}_K|{\bx_2}) - h({\bf r}_1,{\bf r}_2,\ldots ,{\bf r}_K|{\bx_1 , \bx_2}).
\]
From (\ref{relayrx}),
\[{\bf r}_k = \sqrt{\frac{P}{M}}{\bf H}_k{\bx_1} +  \sqrt{\frac{P}{M}}{\bf G}^{r}_k{\bx_2} + {\bf n}_k.\]
%Moreover
%\[ I({\bx_1};{\bf r}_1,{\bf r}_2,\ldots ,{\bf r}_K, {\bx_2})  =
%I\left({\bx_1};\frac{{\bf r}_1}{\sqrt{K}},\frac{{\bf r}_2}{\sqrt{K}},\ldots ,\frac{{\bf r}_K}{\sqrt{K}}, {\bx_2}\right)\]
Since ${\bf G}_k^{r}$ is known at relay $k$,
\[h\left({\bf r}_1,{\bf r}_2,\ldots ,{\bf r}_K|{\bx_2}\right) =
h\left(\sqrt{\frac{P}{M}} {\bf H}_1{\bx_1} + {\bf n}_1,
\sqrt{\frac{P}{M}} {\bf H}_2{\bx_1} + {\bf n}_2,
\ldots,
\sqrt{\frac{P}{M}}{\bf H}_K{\bx_1} + {\bf n}_K|{\bx_2}\right).\]
Since conditioning can only decrease entropy,
\[h\left({\bf r}_1,{\bf r}_2,\ldots ,{\bf r}_K|{\bx_2}\right) \le
h\left(\sqrt{\frac{P}{M}} {\bf H}_1{\bx_1} + {\bf n}_1,
\sqrt{\frac{P}{M}} {\bf H}_2{\bx_1} + {\bf n}_2,
\ldots,
\sqrt{\frac{P}{M}}{\bf H}_K{\bx_1} + {\bf n}_K\right).\]

With perfect knowledge of ${\bf H}_k$ and ${\bf G}_k^{r}$ at relay $k$,
\[h\left({\bf r}_1,{\bf r}_2,\ldots ,{\bf r}_K|{\bx_1 , \bx_2}\right)
= h\left({\bf n}_1, {\bf n}_2,\ldots , {\bf n}_K\right),\]
and it follows that
\begin{eqnarray}\nonumber
I({\bx_1};{\bf r}_1,{\bf r}_2,\ldots ,{\bf r}_K|{\bx_2}) & \le &
h\left(\sqrt{\frac{P}{M}}{\bf H}_1{\bx_1} + {\bf n}_1,
\sqrt{\frac{P}{M}}{\bf H}_2{\bx_1} + {\bf n}_2,
\ldots,
\sqrt{\frac{P}{M}}{\bf H}_K{\bx_1} + {\bf n}_K\right)\\\label{itdummy3}
& & -  h\left({\bf n}_1,{\bf n}_2,\ldots ,{\bf n}_K\right).
\end{eqnarray}
Thus, we have shown that $R_{12}$ is upper bounded by the maximum rate
from $T_1$ to $r_1, \ldots, r_K$ without any interference from $T_2$ and
when all $r_k$'s can collaborate to decode the message, which is quite
intuitive.
Using results from \cite{Telatar1999}
when CSI is known only at the receiver, the R.H.S. of (\ref{itdummy3}) is
upper bounded by
$\log\det\left({\bf I}_M + \sum_{k=1}^{K}\frac{P}{M}{\bf H}_k^*{\bf H}_k\right)$, which implies
\begin{equation}
I({\bx_1};{\bf r}_1,{\bf r}_2,\ldots ,{\bf r}_K|{\bx_2}) \le \alpha\log\det\left({\bf I}_M + \sum_{k=1}^{K}\frac{P}{M}{\bf H}_k^*{\bf H}_k\right)
\end{equation} and therefore, from (\ref{alphaupbound1})
\begin{equation}
\label{upboundbc1}R_{12}\le \alpha I({\bx_1};{\bf r}_1,{\bf r}_2,\ldots ,{\bf r}_K|{\bx_2}) \le \alpha\log\det\left({\bf I}_M + \sum_{k=1}^{K}\frac{P}{M}{\bf H}_k^*{\bf H}_k\right).\end{equation}
Interchanging the roles of ${\bx_1}$ and ${\bx_2}$ and replacing
${\bf H}_k$ with ${\bf G}_k$,
\begin{equation}
\label{upboundbc2}R_{21}\le \alpha I({\bx_2};{\bf r}_1,{\bf r}_2,\ldots ,{\bf r}_K|{\bx_1}) \le \alpha\log\det\left({\bf I}_M + \sum_{k=1}^{K}\frac{P}{KM}{\bf G}_k^{r*}{\bf G}_k^{r}\right).
\end{equation}

{\bf Multiple access cut} - Again by using the cutset bound, we bound the
maximum rate of information transfer $R_{12}$ ($R_{21}$) from $T_1
\rightarrow T_2$ ($T_1 \rightarrow T_2$) by the maximum rate of
information transfer across the multiple access cut as shown in Fig.
\ref{cutsetmac}. Using cutset bound, $R_{12}$ and $R_{21}$ are bounded by
\begin{equation}
\label{initialupboundmac1}
R_{12} \le (1-\alpha)I({\bx_1},  {\bf t}_1,  {\bf t}_2,  \ldots, {\bf t}_K ;  \by_2 | \bx_2)
\end{equation}
\begin{equation}
\label{upboundmac2}
R_{21} \le (1-\alpha)I({\bx_2}, {\bf t}_1, {\bf t}_2, \ldots,  {\bf t}_K ; \by_1 | \bx_1).
\end{equation}
Now,
\begin{eqnarray*}
I({\bx_1}, {\bf t}_1, {\bf t}_2, \ldots,  {\bf t}_K ; \by_2 | \bx_2)
%& = &
%I({\bf t}_1, {\bf t}_2, \ldots, {\bf t}_K ; \by_1 | \bx_2) \\
%& & + I({\bx_1} ; \by_1 | {\bf t}_1, {\bf t}_2, \ldots, {\bf t}_K, \bx_2)\\
&=&h(\by_2 | {\bx_2}) - h(\by_2| {\bf t}_1, {\bf t}_2, \ldots, {\bf t}_K, \bx_2) \\
& & + h(\by_2| {\bf t}_1, {\bf t}_2, \ldots, {\bf t}_K, {\bx_2})
- h(\by_2| {\bf t}_1, {\bf t}_2, \ldots, {\bf t}_K,  {\bx_1}, \bx_2). \\
\end{eqnarray*}
Note that given ${\bf t}_1, {\bf t}_2, \ldots, {\bf t}_K$, $\by_2$ is
independent of ${\bx_1}$ and ${\bx_2}$,
\[h(\by_2| {\bf t}_1, {\bf t}_2, \ldots, {\bf t}_K,  {\bx_1}, {\bx_2}) =
 h(\by_2| {\bf t}_1, {\bf t}_2, \ldots, {\bf t}_K, {\bx_2}) = h(\by_2| {\bf t}_1, {\bf t}_2, \ldots, {\bf t}_K).\]
Therefore
\[I({\bx_1}, {\bf t}_1, {\bf t}_2, \ldots,  {\bf t}_K ; \by_2 | {\bx_2})
= h(\by_2 |{\bx_2}) -  h(\by_2| {\bf t}_1, {\bf t}_2, \ldots, {\bf t}_K). \]
Since conditioning can only reduce entropy,
\[I({\bx_1}, {\bf t}_1, {\bf t}_2, \ldots,  {\bf t}_K ; \by_2 | {\bx_2})
\le  h(\by_2) -  h(\by_2| {\bf t}_1, {\bf t}_2, \ldots, {\bf t}_K), \] and
by definition of mutual information
\[I({\bx_1}, {\bf t}_1, {\bf t}_2, \ldots,  {\bf t}_K ; \by_2 | {\bx_2})
\le  I({\bf t}_1, {\bf t}_2, \ldots, {\bf t}_K, \by_2). \]
Hence from (\ref{initialupboundmac1}),
\begin{equation}
\label{t1t2mac}
R_{12} \le (1-\alpha)I({\bf t}_1, {\bf t}_2, \ldots, {\bf t}_K ; \by_2).
\end{equation}
Following similar steps we can also bound $R_{21}$ as,
\begin{equation}
\label{t2t1mac}
R_{21} \le (1-\alpha)I({\bf t}_1, {\bf t}_2, \ldots, {\bf t}_K ; \by_1).
\end{equation}
Thus, $R_{12}, R_{21}$ are bounded by the maximum rate of
information from $r_1, \ldots, r_K$ to $T_1$ or $T_2$. Next, we
compute the maximum rate of
information from $r_1, \ldots, r_K$ to $T_1$ or $T_2$.
Recall from (\ref {t2rx}) that the received signal $\by_2$ is
given by
\[\by_2 = \sum_{k=1}^{K}{\bf G}_k{\bf t}_k + {\bf z}_2.\]
Note that \[I({\bf t}_1, {\bf t}_2, \ldots, {\bf t}_K; \by_2) =
I\left({\bf t}_1, {\bf t}_2, \ldots, {\bf t}_K; \frac{\by_2}{\sqrt{K}}\right).\]
Dividing $\by_2$ by $\sqrt{K}$, we get
\[\frac{\by_2}{\sqrt{K}} = \frac{1}{\sqrt{K}}\sum_{k=1}^{K}{\bf G}_k{\bf t}_k + \frac{{\bf z}_2}{\sqrt{K}}.\]
This can also be written as
\[\frac{\by_2}{\sqrt{K}} = \underbrace{\frac{1}{\sqrt{K}}
\left[{\bf G}_1 \ {\bf G}_2 \ \ldots \
{\bf G}_K\right]}_{\Phi}
\left[{\bf t}_1 {\bf t}_2 \ldots {\bf t}_K\right]^T + \frac{{\bf z}_2}{\sqrt{K}}.\]
Note that $\Phi$ is a $M \times NK$ matrix.
Now assuming that all the relays know ${\bf G}_k, \ \forall
 k$ (allowing cooperation among all relays), with sum power available across all relays bounded by $P_R$, we have
from \cite{Telatar1999},
\begin{equation}
\label{upboundmac1}
R_{12}\le (1-\alpha)I\left({\bf t}_1, {\bf t}_2, \ldots, {\bf t}_K ;\frac{\by_2}{\sqrt{K}}\right) \le (1-\alpha)\sum_{l=1}^{\min{\{NK,M\}}}
\max{\left\{0,\log\left(K\lambda_l\nu\right)\right\}}
%\max{\left\{0, \log{\left( \frac{K\lambda_l\nu}{\}\right) }\right\}
\end{equation}
where $\lambda_l, l=1,2, \ldots, \min{\{NK,M\}}$ are the eigen values of $\Phi\Phi^*$ matrix  and $\nu$ is chosen such that
\[ \sum_{l=1}^{\min{\{NK,M\}}} \max{\{0, \nu - \frac{1}{\lambda_l}\}} = P_R. \]
Similarly, one can obtain the bound for $R_{21}$ by replacing $\bG_k$ by
$\bH_k^{r}$.

Combining (\ref{upboundbc1}), (\ref{upboundbc2}) and (\ref{upboundmac1}), gives the  upper bound on the capacity region
of the two-way relay channel.
Comparing the upper bound with the lower bound obtained
using the dual channel matching (\ref{lowerbddcm1},\ref{lowerbddcm2}), one can see that they do not match for
any arbitrary value of $K$.
In the asymptotic regime, however, they can
be shown to be only an ${\cal O}(1)$ term away as $K\rightarrow \infty$, as proved in
the next Theorem. This asymptotic result implies two things, one that the
performance of the dual channel matching, and consequently the optimal AF
strategy (which we don't know for $M>1$), does not degrade in comparison to the upper bound with increasing $K$, and two, it provides us with
the capacity scaling law of the two-way relay channel.

\begin{figure}
\centering
\includegraphics[height= 3in]{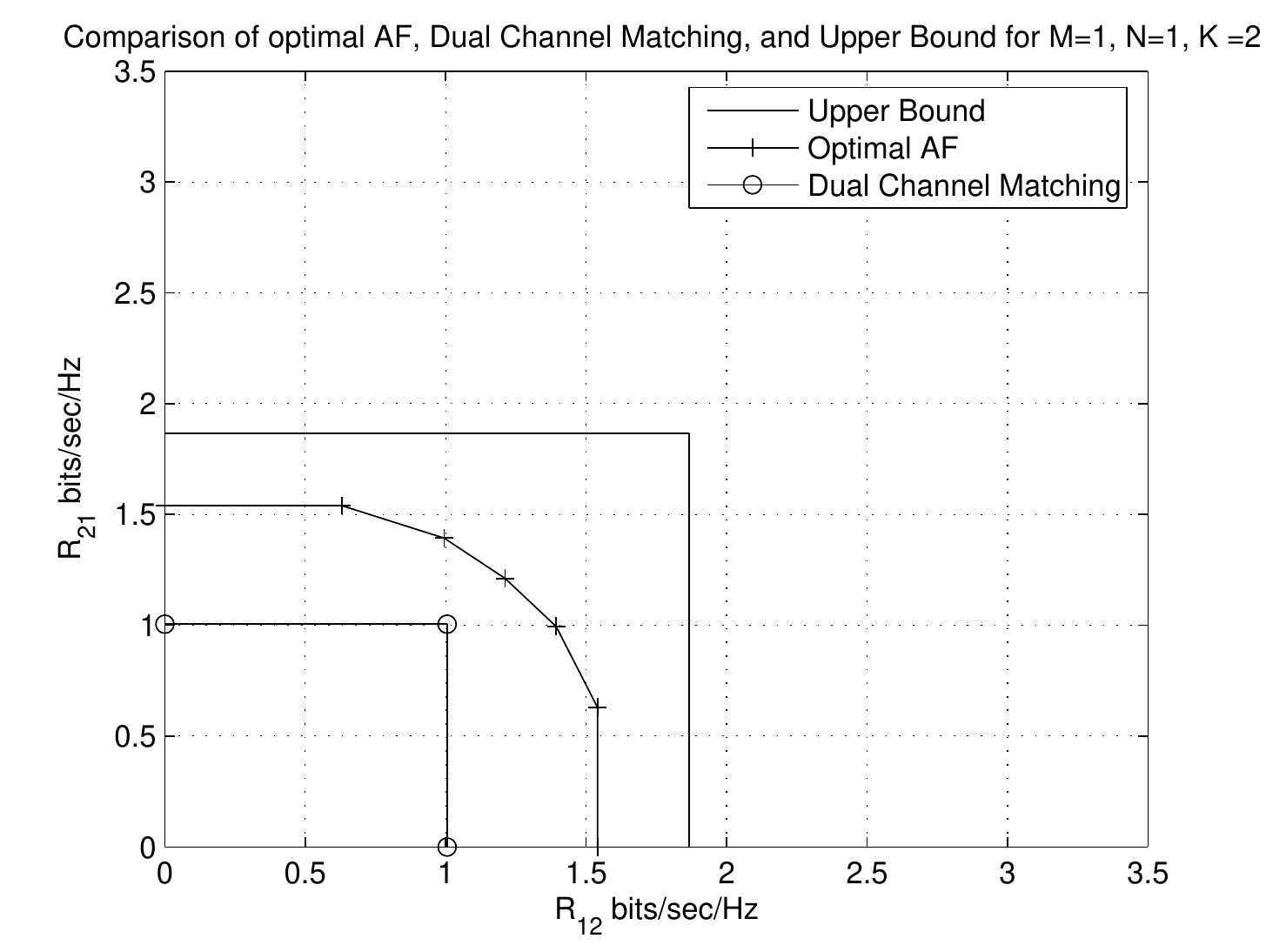}
\caption{Comparison of upper and lower bound of the capacity region of the two-way relay channel with $K=2, M=1,N=1,$ $P=P_R=10dB$}
\label{capcomp1}
\end{figure}

\begin{figure}
\centering
\includegraphics[height= 3in]{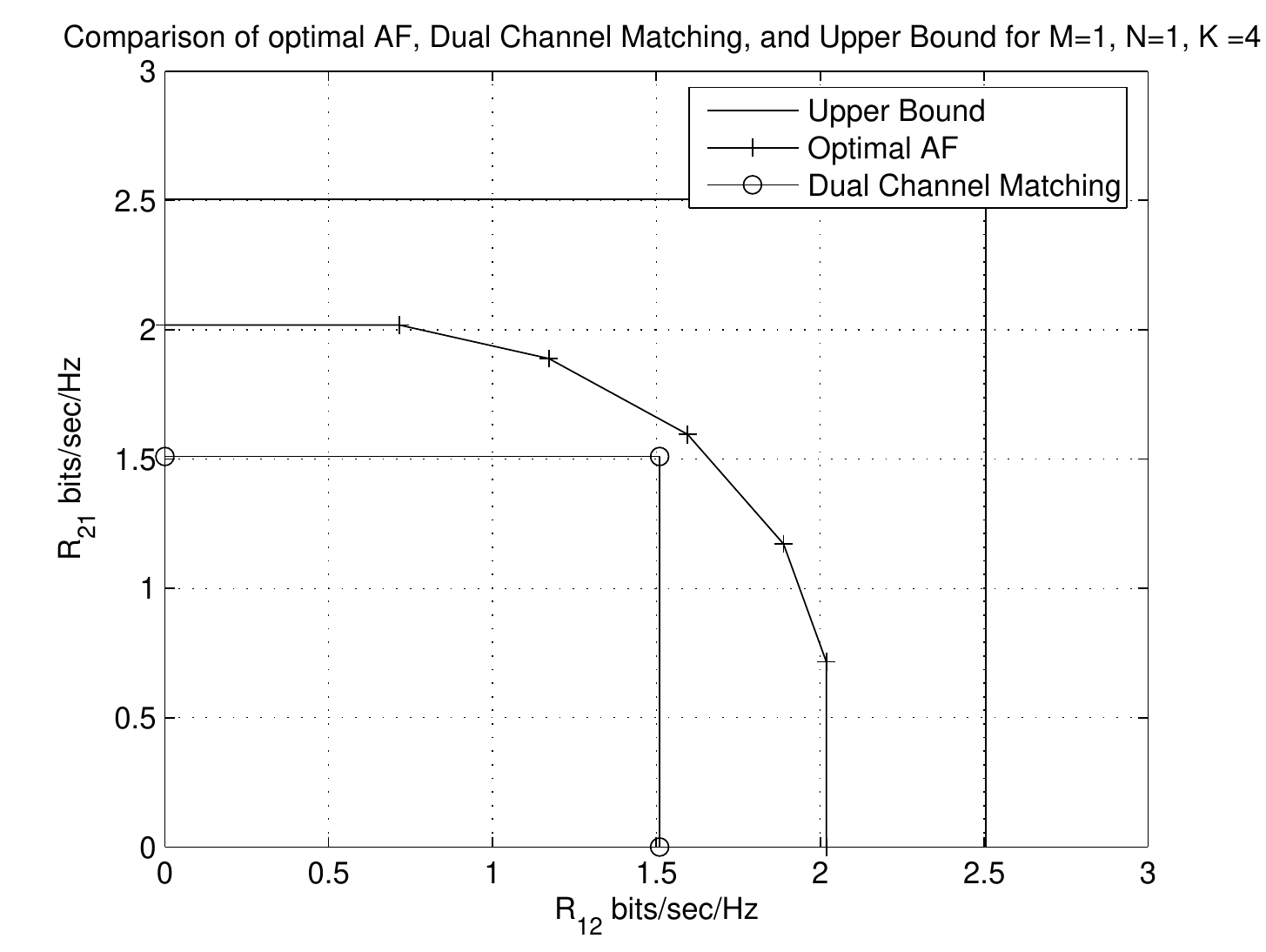}
\caption{Comparison of upper and lower bound of the capacity region of the two-way relay channel
with $K=4, M=1,N=1$, $P=P_R=10dB$}
\label{capcomp2}
\end{figure}

In Figs. \ref{capcomp1} and \ref{capcomp2}, we plot the
achievable rate region of the optimal AF strategy, the lower bound obtained using
dual channel matching, and the upper bound for $K=2$ and $K=4$, with $M=1, \ N=1$ and $P=P_R =10dB$ with sum rate constraint across relays.
Note that the achievable rate region of the optimal AF region is symmetric,
as expected, because of the symmetry in parameters of communication in both
directions in a two-way relay channel. Another important point to note here is that, the
 achievable rates of dual channel matching are quite close to that of the
optimal AF strategy, even though it uses only local CSI. Thus, dual channel
matching is a good candidate for AF in practical implementation of two-way relay channels.
Also notice that the difference between the  upper and lower bound is less than the  $3$ bit
 bound of \cite{Tse2008}.

Next, we prove that the lower bound (dual channel matching) and the upper bound 
on the achievable 
rate region of the two-way relay channel are only an
${\cal O}(1)$ as $K\rightarrow \infty$. We prove the theorem by approximating
the upper bound in the $K\rightarrow \infty$ and comparing it with the
asymptotic lower bound obtained in (\ref{lbdcmass1}, \ref{lbdcmass2}).
\begin{thm}
\label{asympcap}The upper and lower bounds on the capacity region of the two-way
relay channel differ by a ${\cal O}(1)$ term as $K\rightarrow \infty$, and the
capacity scaling law is given by
\begin{eqnarray*}
R_{12} &\le &  \frac{M}{2} \log K + {\cal O}(1),\\
R_{12} &\le &  \frac{M}{2} \log K + {\cal O}(1).
\end{eqnarray*}
\end{thm}
\begin{proof}
We first approximate the broadcast cut upper bound (\ref{upboundbc1}) as
$K\rightarrow \infty$.
From (\ref{upboundbc1})
\begin{equation}R_{12}\le \alpha I({\bx_1};{\bf r}_1,{\bf r}_2,\ldots ,{\bf r}_K|{\bx_2}) \le \alpha\log\det\left({\bf I}_M + \sum_{k=1}^{K}\frac{P}{M}{\bf H}_k^*{\bf H}_k\right).\end{equation}
Consider
\[\log\det\left({\bf I}_M + \sum_{k=1}^{K}\frac{P}{M}{\bf H}_k^*{\bf H}_k\right) - \log \det K\bI_M = \log\det\left(\frac{1}{K}{\bf I}_M + \frac{1}{K}\sum_{k=1}^{K}\frac{P}{M}{\bf H}_k^*{\bf H}_k\right).\]
Using strong law of large numbers
\[\lim_{K \rightarrow \infty}\frac{1}{K}\sum_{k=1}^{K}\frac{P}{M}{\bf H}_k^*{\bf H}_k \xrightarrow{w.p.1}\frac{PN}{M}\bI_{M}, \
 \text{since} \ {\bbE}\{{\bf H}_k^*{\bf H}_k\} = N{\bf I}_M,\]
and it follows that
\[\log\det\left(\frac{1}{K}{\bf I}_M + \frac{1}{K}\sum_{k=1}^{K}\frac{P}{M}{\bf H}_k^*{\bf H}_k\right) \rightarrow M\log\left(\frac{PN}{M}\right),\]
which using (\ref{upboundbc1}) implies
\begin{equation}
\label{upboundassbc1}
\lim_{K \rightarrow \infty}R_{12} \mapright{w.p. 1}  \alpha M\log K + {\cal O}(1),
\end{equation}
since $M, N, P$ are finite integers. Similarly,
\begin{equation}
\label{upboundassbc2}
\lim_{K \rightarrow \infty}R_{21} \mapright{w.p. 1} \alpha M\log K + {\cal O}(1).
\end{equation}

Next, we approximate the upper bound of the multiple access cut. From
(\ref{upboundmac1}),
\begin{equation}
\label{upboundmacdummy}
R_{12}\le (1-\alpha)I\left({\bf t}_1, {\bf t}_2, \ldots, {\bf t}_K ;\frac{\by_2}{\sqrt{K}}\right) \le (1-\alpha)\sum_{l=1}^{\min{\{NK,M\}}}
\max{\left\{0,\log\left(K\lambda_l\nu\right)\right\}},
%\max{\left\{0, \log{\left( \frac{K\lambda_l\nu}{\}\right) }\right\}
\end{equation}
where $\lambda_l, l=1,2, \ldots, \min{\{NK,M\}}$ are the eigen values of $\Phi\Phi^*$ matrix  and $\nu$ is chosen such that
\[ \sum_{l=1}^{\min{\{NK,
M\}}} \max{\{0, \nu - \frac{1}{\lambda_l}\}} = P_R. \]

By definition
$\Phi\Phi^* = \frac{1}{K}\sum_{k=1}^K \bG_k\bG_k^*$.
From strong law of large numbers
\[\lim_{K \rightarrow \infty}\frac{1}{K}\sum_{k=1}^K\bG_k\bG_k^* \xrightarrow{w.p. 1} N{\bf I}_M.\]
 Therefore
\[\lambda_i = N  \ \ \forall \ i = 1,2, \ldots M, \implies \nu = \left(\frac{P_R}{M} + \frac{1}{N}\right),\] and from (\ref{upboundmacdummy})
\[R_{12} \mapright{w.p. 1} (1-\alpha)\sum_{l=1}^{M} \log\left(KN\left(\frac{P_R}{M} + \frac{1}{N}\right)\right),\] and
consequently, as $K \rightarrow \infty$
\begin{equation}
\label{upboundassmac1}
R_{12} \mapright{w.p. 1} (1-\alpha)M\log K +{\cal O}(1),
\end{equation} and similarly
\begin{equation}
\label{upboundassmac2}R_{21} \mapright{w.p. 1} (1-\alpha)M\log K +{\cal O}(1).\end{equation}
Combining (\ref{upboundassbc1},\ref{upboundassbc2}) and (\ref{upboundassmac1},\ref{upboundassmac2})
\begin{eqnarray}\nonumber
R_{12} &\le & \min\{\alpha, 1-\alpha\} M \log K + {\cal O}(1) \le
\frac{M}{2} \log K + {\cal O}(1),\\\label{asymupbound}
R_{21} &\le & \min\{\alpha, 1-\alpha\} M \log K + {\cal O}(1) \le
\frac{M}{2} \log K + {\cal O}(1).
\end{eqnarray}

Comparing (\ref{asymupbound}) to the asymptotic lower bound
(\ref{lbdcmass1}, \ref{lbdcmass2}) we conclude that
(a) upper and lower bounds on the capacity region of the two-way
relay channel differ by a ${\cal O}(1)$ term as $K\rightarrow \infty$, and
(b) the capacity scaling law is given by
\begin{eqnarray*}
R_{12} &\le &  \frac{M}{2} \log K + {\cal O}(1),\\
R_{21} &\le & \frac{M}{2} \log K + {\cal O}(1).
\end{eqnarray*}
\end{proof}

\begin{figure}
\centering
\includegraphics[height= 3in]{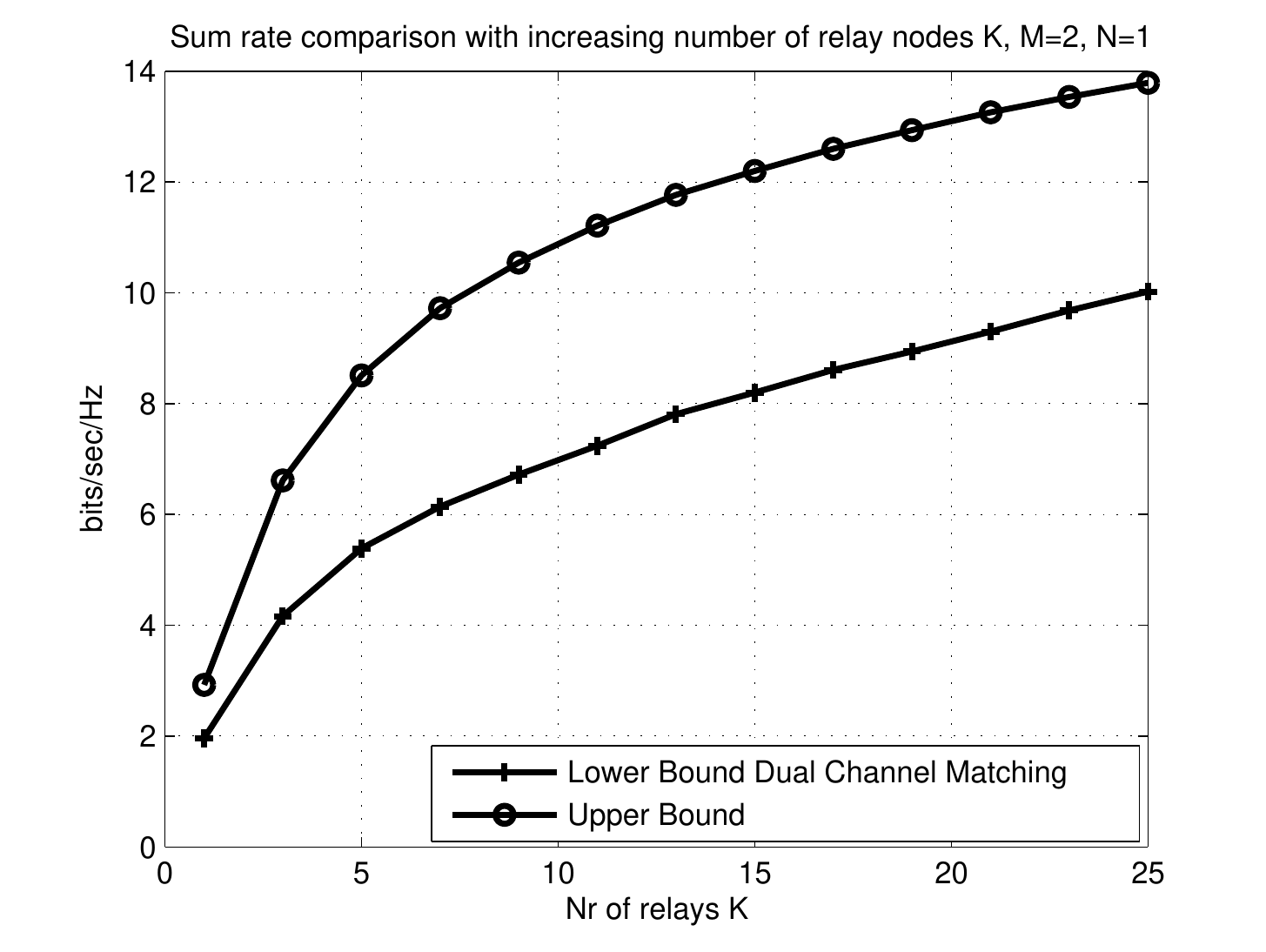}
\caption{Capacity scaling of two-way relay channel with $M=2, N =1, P=P_R=10dB$.}
\label{capplot}
\end{figure}

To illustrate the result of Theorem \ref{asympcap}, in Fig.\ref{capplot}, we compare the lower (dual channel matching) and upper bound on the sum rate $R_{12}+R_{21}$, and show that they both scale
similarly with increasing $K$ for $M=2, N =1$, $P=P_R=10dB$ with sum rate constraint across relays.

{\it Discussion:}
In this section we obtained upper bounds on the capacity region  of the two-way relay channel, and compared it with the dual channel matching lower bound.
To compute the upper bound we used the cut-set
bound and the capacity results of \cite{Telatar1999}. The lower and upper bound
expressions do not match in general, however, in the asymptotic case, where the number of
relays are large, $K \rightarrow \infty$, we showed that they are only an
${\cal O }(1)$ term away from each other. Thus, the dual channel matching and
consequently, the optimal AF strategy are almost optimal in the asymptotic
regime.
For the finite number of relay nodes (finite $K$), we use Monte Carlo
simulations to quantify the gap between the lower and the upper bound. From
Figs. \ref{capcomp1} and \ref{capcomp2}, we can see that gap between the lower (dual channel matching)
and upper bound is rather small, and inside the $3$ bit bound of \cite{Tse2008}.

Another important observation to make is that the lower bound
with dual channel matching was obtained using $\alpha = \frac{1}{2}$ i.e.
$T_1$ and $T_2$ transmit and receive for equal amount of time.
Since this lower bound is only a ${\cal O}(1)$ term away
from the upper bound (arbitrary $\alpha$), distributing equal amount of
time for transmit and receive phase is optimal in achieving the right
capacity scaling.

Compared to the asymptotic results on the one-way relay channel
\cite{Bolcskei2006, Gastpar2002b}, our results show that by  two-way relay channel one can remove
the $\frac{1}{2}$ rate loss factor on the capacity, which comes from the
half-duplex assumption on the terminals and relays.
Therefore with two-way relay channel one can achieve unidirectional
full-duplex performance with half-duplex terminals.

%XXX CHECK MY CHANGES BELOW. STILL CONFUSING WITH ALL THE RESPECTIVELYS. MAYBE WE CAN TALK ABOUT T1->T2 FOR SIMPLICTY OF EXPLANATION THEN SUMMARIZE THE REVERSE?
%Rahul: Hopefully its clear now

%XXX THIS DOESN'T MAKE SENSE - DO YOU MEAN WE ASSUME THE RELAY KNOWS THESE THINGS? DOESN'T MAKE SENSE WITH THE ABOVE
%Rahul Hope its clear now

\section{Diversity-Multiplexing Tradeoff}
\label{sec:dmt}
In this section we consider a two-way relay channel with a single relay
node, and characterize its DM-tradeoff. We consider both the full-duplex and
half-duplex nodes, where $T_1$ and $T_2$ have $m_1$ and $m_2$ antennas,
respectively, and the relay node has $m_r$ antennas. An important
difference in this section from the previous ones is
the presence of direct link between $T_1$ and $T_2$ as shown in Fig.
\ref{blkdmt}.

To characterize the DM-tradeoff, for both the full-duplex and half-duplex case,  we first obtain an upper bound on the DM-tradeoff and then propose a modified
 CF strategy to achieve the upper bound. We first discuss the
full-duplex case followed by the half-duplex case.

\subsection{DM-tradeoff of Full-Duplex Two-Way Relay Channel}
\label{subsec:dmtfd}
The signal model for this section is as follows. Let $\bx_1$, $\bx_2$ and
$\bx_r$ be the signal transmitted from $T_1$, $T_2$ and the relay node,
respectively. Similarly, Let $\by_1$, $\by_2$ and
$\by_r$ be the signal received at $T_1$, $T_2$ and the relay node,
respectively. Recall that channel coefficient between $T_1$ and relay node is
denoted by $\bH$, between $T_1$ and $T_2$ is denoted by $\bH_{12}$,
between the relay node and $T_2$ is denoted by $\bG$, where note that, compared to previous sections, we have dropped the subscript index of relay node, since we only consider one relay. All the channel
coefficients in the reverse direction (right to left) are denoted by
channel coefficient in the forward direction (left to right) with an added
superscript $r$, e.g. the channel coefficients between the relay node and
$T_1$ is denoted by $\bH^r$. Let the transmit power at $T_1$, $T_2$ and the
relay node be $P$\footnote{Having different transmit power constraints for
$T_1$, $T_2$ and the relay node do not change the DM-tradeoff.}.
Then,
\begin{eqnarray}
\nonumber
\by_1 &=& \sqrt\frac{{P}}{m_2}\bH_{12}^r\bx_2 + \sqrt\frac{{P}}{m_r}\bH^r\bx_r
+ \bn_1, \\\nonumber
\by_2 &=& \sqrt\frac{{P}}{m_1}\bH_{12}\bx_1 + \sqrt\frac{{P}}{m_r}\bG\bx_r
+ \bn_2, \\\label{rxsigdmtfd}
\by_r &=& \sqrt\frac{{P}}{m_1}\bH\bx_1 + \sqrt\frac{{P}}{m_2}\bG^r\bx_2
+ \bn_r.
\end{eqnarray}
Let the rate of transmission from $T_1$ to $T_2$ and $T_2$ to $T_1$ be $R_{12}$
and $R_{21}$, respectively.
Following \cite{Zheng2003}, let ${\cal C}_{12}(\SNR)$ and ${\cal C}_{21}(\SNR)$
be the family of codes, one for each $\SNR$ for transmission from $T_1$ to
$T_2$, and $T_2$ to $T_1$, respectively.
Then we define $r_{12}$ ($r_{21}$ similarly) as the multiplexing gain of
${\cal C}_{12}(\SNR)$ if the data rate $R_{12}(\SNR)$ ( $R_{21}(\SNR)$)
of ${\cal C}_{12}(\SNR)$ (${\cal C}_{21}(\SNR)$)
scales as $r_{12}$ ($r_{21}$) with respect to $\log \SNR$, i.e.
\[\lim_{\SNR\rightarrow \infty}\frac{R_{12}(\SNR)}{\log \SNR} = r_{12}\]
and $d_{12}(r_{12}, r_{21})$ ($d_{21}(r_{12}, r_{21})$) as the rate of fall of
probability of error
$P_{e12}$ ($P_{e21}$) of ${\cal C}_{12}(\SNR)$ (${\cal C}_{21}(\SNR)$)
with respect to \SNR, i.e.
\[P_{e12}(\SNR) \expeq \SNR^{-d_{12}(r_{12}, r_{21})}.\]
The negative of the $\SNR$ exponent of the error probability
$d_{12}(r_{12}, r_{21})$ or $d_{21}(r_{12}, r_{21})$ captures the DM-tradeoff,
where $d_{12}(r_{12}, r_{21})$ ($d_{21}(r_{12}, r_{21})$) is the maximum
diversity gain possible from $T_1$ to $T_2$ ($T_2$ to $T_1$) for a given
$r_{12}$ and $r_{21}$.
Note that the error probability $P_{e12}(\SNR)$ and $P_{e21}(\SNR)$ are
functions of both $r_{12}$ and $r_{21}$ because of simultaneous transmission
between $T_1$ and $T_2$.

Next, we upper bound the DM-tradeoff of the two-way relay channel, the region
spanned by $d_{12}(r_{12}, r_{21})$ and $d_{21}(r_{12}, r_{21})$, by allowing
cooperation between $T_1$ and relay, and $T_2$ and relay node.

\begin{lemma}
\label{dmtupboundfd} The DM-tradeoff of a two-way relay channel is upper
bounded by
\begin{eqnarray*}
d_{12}(r_{12}, r_{21}) &\le& \min\{(m_1-r_{12})(m_r+m_2-r_{12}), \
(m_1+m_r-r_{12})(m_2-r_{12})\}, \\
d_{21}(r_{12}, r_{21}) &\le& \min\{(m_2-r_{21})(m_r+m_1-r_{21}), \
(m_2+m_r-r_{21})(m_1-r_{21})\},  \ \forall r_{12}, \ r_{21}.
\end{eqnarray*}
\end{lemma}
\begin{proof} We will prove the lemma only for $d_{12}(r_{12}, r_{21})$. For
$d_{21}(r_{12}, r_{21})$ it follows similarly. Consider the case when $T_2$ has
no data to send to $T_1$. This assumption can only improve
$d_{12}(r_{12}, r_{21})$. Then first assume that the relay node and $T_2$
are co-located and can cooperate perfectly. In this case, the
communication model from $T_1$ to $T_2$ is a point to point MIMO
channel with $m_1$ transmit antennas and $m_r+m_2$ receive antennas. The
DM-tradeoff of this MIMO channel is $(m_1-r_{12})(m_r+m_2-r_{12})$, and
since this point to point MIMO channel is better than our original two-way
relay channel, $d_{12}(r_{12}, r_{21}) \le (m_1-r_{12})(m_r+m_2-r_{12})$
\footnote{This upper bound is valid as long as the coherence time $T_c$ is
smaller than the time it takes for $T_2$ to compute the channel
coefficients and feed them back to $T_1$, which is at least $m_1+m_2$
\cite{Hassibi2003}. Otherwise, $T_2$ can help $T_1$ in acquiring transmit
CSI, for which case, potentially infinite diversity gain can be achieved
\cite{Biglieri2001},
violating the present upper bound.} .
Next, we assume that $T_1$ is co-located with relay node and both of them can
perfectly cooperate for transmission to $T_2$.
This setting is equivalent to a MIMO channel with $m_1+m_r$ transmit and
$m_2$ receive antenna with DM-tradeoff $(m_1+m_r-r_{12})(m_2-r_{12})$. Again,
this point to point MIMO channel is better than our original two-way relay
channel and hence $d_{12}(r_{12}, r_{21}) \le (m_1+m_r-r_{12})(m_2-r_{12})$,
which completes the proof.
\end{proof}

To achieve this upper bound we consider the CF strategy \cite{Cover1979}, with a slight
modification and prove that it is sufficient, to achieve the optimal DM-tradeoff.
We make few changes to the original CF strategy \cite{Cover1979}
to suit the two-way relay channel communication, which are as follows.
Let the rate of transmission from $T_1$ to $T_2$ and $T_2$ to $T_1$ be $R_{12}$
and $R_{21}$, respectively. Instead of generating only one codebook at $T_1$
as in \cite{Cover1979}, both $T_1$ and $T_2$ generate $2^{nR_{12}}$ and $2^{nR_{21}}$
independent
and identically distributed $x_1^n$ and $x_2^n$ according to distribution
$p(x_1^n) = \prod_{i=1}^np(x_{1i})$ and $p(x_2^n) = \prod_{i=1}^np(x_{2i})$,
respectively. The codebook generation at the relay and the relay compression
and transmission remains the same as in \cite{Cover1979}, i.e.
the relay node generates $2^{nR_{0}}$ independent and identically distributed
$x_r^n$ according to distribution $p(x_r^n) = \prod_{i=1}^np(x_{ri})$ and label
them $x_r(s), \ s\in [1, 2^{nR_0}]$, and for each $x_r(s)$ generates $2^{n{\hat R}}$ ${\hat y}$'s, each with probability $p({\hat y}|x_r(s)) = \prod_{i=1}^np({\hat y}_{i}|x_{ri}(s)) $.  Label these ${\hat y}(z|s), s \in [1,2^{nR_0}]$ and $z \in [1,2^{n{\hat R}}]$ and randomly partition the set $ [1,2^{n{\hat R}}]$ into
$2^{nR_0}$ cells $S_{s}, \ s\in [1, 2^{nR_0}]$. Let in block $i$ the message to send from $T_1$ is $w_i$, and from $T_2$ is $v_i$, then $T_1$ sends
$x_1(w_i)$, $T_2$ sends $x_2(v_i)$ and the relay sends $x_r(s_i)$
if $z_i \in s_i$, where ${\hat y}(z_i|s_{i-1}), y_r(i-1), x_r(s_{i-1})$
are jointly typical.
Decoding at both $T_1$ and $T_2$ remains the same as in \cite{Cover1979}, however, note that in this case $T_1$ knows $x_1(w_i)$ and $T_2$ knows $x_2(v_i)$ apriori and therefore can use them to decode $v_i$ and $w_i$ respectively. This
strategy has been previously considered in \cite{Rankov2006} to obtain achievable rate
region.

With this two-way CF strategy, the following rates are
achievable,
\begin{eqnarray}
\nonumber
R_{12} &\le& I(\bx_1;\by_2 \hat{\by} | \bx_r \bx_2), \\
\label{cfratefd}
R_{21} &\le& I(\bx_2;\by_1 \hat{\by} | \bx_r \bx_1),
\end{eqnarray}
with the compression rate constraint
\begin{eqnarray}
\label{compconstraintfd}
\max\{I(\by_r; \hat{\by} | \bx_r \bx_1 \by_1), I(\by_r; \hat{\by} | \bx_r \bx_2 \by_2)\} \le
\min\{I(\bx_r;\by_1 | \bx_1), I(\bx_r;\by_2 | \bx_r \bx_2)\}.
\end{eqnarray}
The rate region and the compression constraint are a little different from
\cite{Cover1979}. The rate region differs because of conditioning
by $\bx_1$ or $\bx_2$, which is due to the prior knowledge of $\bx_1$ at $T_1$ and
$\bx_2$ at $T_2$. The new compression rate constraint incorporates the
condition that the quantized version of $\by_r$, $\hat{\by}_r$ can be decoded
at both $T_1$ and $T_2$. In the next Theorem
we compute the outage exponents for (\ref{cfratefd}) and show that they match with the
exponents of the upper bound.

\begin{thm} CF strategy achieves the DM-tradeoff upper bound
(Lemma \ref{dmtupboundfd}).
\end{thm}
\begin{proof}
To prove the Theorem we will compute the achievable DM-tradeoff of the CF
strategy (\ref{cfratefd}) and show that it matches with the upper bound.

To compute the achievable rates subject to the compression rate
constraints for the signal model (\ref{rxsigdmtfd}), we fix ${\hat
\by} = \by_r + \bn_q$, where $\bn_q$ is $m_r \times 1$ vector with
covariance matrix ${\hat N}\bI_{m_r}$. Also, we choose $\bx_1$,
$\bx_2$, and $\bx_r$ to be  complex Gaussian with covariance
matrices $\frac{P}{m_1}\bI_{m_1}$,  $\frac{P}{m_2}\bI_{m_2}$, and
$\frac{P}{m_r}\bI_{m_r}$, and independent of each other.
respectively. Next, we compute the various mutual information
expressions to derive the achievable DM-tradeoff of the CF strategy.
By the definition of the mutual information
\begin{eqnarray*}
I(\bx_1;\by_2 \hat{\by} | \bx_r \bx_2) & = & h(\by_2 \hat{\by} | \bx_r \bx_2) - h(\by_2 \hat{\by} | \bx_r \bx_2 \bx_1). \end{eqnarray*}
From (\ref{rxsigdmtfd}), arranging $\by_2 \ \hat{\by}$ in a vectorized form
we get
\begin{equation}
\label{vecrxsigdmt}
\left[\begin{array}{c}
\by_2 \\ \hat{\by} \end{array}\right] =
\left[\begin{array}{c}
\sqrt\frac{{P}}{m_1}\bH_{12}\bx_1 + \sqrt\frac{{P}}{m_r}\bG\bx_r + \bn_2, \\
\sqrt\frac{{P}}{m_1}\bH\bx_1 + \sqrt\frac{{P}}{m_2}\bG^r\bx_2 + \bn_r + \bn_q.
\end{array}\right]
\end{equation}
and consequently
\begin{eqnarray}
\label{ents-rd}
h(\by_2 \hat{\by} | \bx_r \bx_2) = \log L^{r2}_{1}, \end{eqnarray}
 where
\begin{equation*}L^{r2}_1 = \det\left(\frac{P}{m_1}\bH_{1}^{r2}\bH_{1}^{r2*} + \left[\begin{array}{cc}\left({\hat N}+1\right)\bI_{m_r} & 0 \\ 0 & \bI_{m_2}\end{array}\right]\right) \ \text{and} \ \bH_{1}^{r2} = [\bH_{12}\ \bH].\end{equation*}
Moreover, from (\ref{vecrxsigdmt})
\begin{eqnarray*}h(\by_2 \hat{\by} | \bx_r \bx_2 \bx_1) = \log \det \left(\left[\begin{array}{cc}(\hat{N}+1)\bI_{m_r} & 0 \\ 0 & \bI_{m_2}\end{array}\right]\right), \end{eqnarray*}
which implies
\begin{eqnarray}
\label{mis-rd}
I(\bx_1;\by_2 \hat{\by} | \bx_r \bx_2) & = & \log\frac{ L_{1}^{r2}}{({\hat N}+1)^{m_r}}.\end{eqnarray}
Similarly, one can show,
\begin{eqnarray*}
I(\bx_2;\by_1 \hat{\by} | \bx_r \bx_1) & = & \log\frac{L_{2}^{r1}}{({\hat N}+1)^{m_r}}, \end{eqnarray*}
where \begin{equation}L^{r1}_2 = \det\left(\frac{P}{m_1}\bH_{2}^{r1}\bH_{2}^{r1*} + \left[\begin{array}{cc}\left({\hat N}+1\right)\bI_{m_r} & 0 \\ 0 & \bI_{m_2}\end{array}\right]\right) \ \text{and} \ \bH_{2}^{r1} = [\bH_{12}^r \ \bG^r]. \end{equation}
Next, we compute the value of ${\hat N}$ that satisfies the
compression rate constraints (\ref{compconstraintfd}).
By the definition of mutual information,
\begin{eqnarray}\nonumber
I(\by_r; \hat{\by} | \bx_r \bx_2 \by_2) &= &h(\hat{\by} | \bx_r \bx_2 \by_2) - h(\hat{\by} | \bx_r \bx_2 \by_2 \by_r) \\\label{micompcon}
& = & h(\hat{\by} \by_2 | \bx_r \bx_2) - h(\by_2 | \bx_r \bx_2) - h(\hat{\by} | \bx_r \bx_2 \by_2 \by_r).\end{eqnarray}
From (\ref{ents-rd}),
$h(\hat{\by} \by_2 | \bx_r \bx_2) = \log L^{r2}_{1}$.
From signal model (\ref{rxsigdmtfd}), it is easy to see that
$h(\by_2 | \bx_r \bx_2) = \log L_{12}$, where
$L_{12} = \det\left(\frac{P}{m_1}\bH_{12}\bH_{12}^* + I_{m_2}\right)$.
Given $\by_r$, $\hat{\by}$ has only the noise term $\bn_q$, and hence
$h(\hat{\by} | \bx_r \bx_1 \by_1 \by_r) = \log {\hat N}^{m_r}$.
Therefore, from (\ref{micompcon}),
\begin{eqnarray}
\label{dmtdummy1}
I(\by_r; \hat{\by} | \bx_r \bx_2 \by_2) = \log \frac{L^{r2}_1}{L_{12}{\hat N}^{m_r}}.\end{eqnarray}
Similarly one can compute
\begin{eqnarray}
\label{dmtdummy2}
I(\by_r; \hat{\by} | \bx_r \bx_1 \by_1) = \log \frac{L^{r1}_2}{L_{21}{\hat N}^{m_r}}, \ \text{where} \ L_{21} = \det\left(\frac{P}{m_2}\bH_{12}^{r}\bH_{12}^{r*} + I_{m_1}\right).\end{eqnarray}
Again using the definition of mutual information,
\begin{eqnarray}\nonumber
I(\bx_r;\by_1|\bx_1) &=& h(\by_1|\bx_1) - h(\by_1|\bx_r \bx_1) \\\label{dmtdummy3}
&=& \log L_{2r}^1 - \log L_{21}, \end{eqnarray}
where $L_{2r}^1 = \det\left(\frac{P}{m_2}\bH_{12}^{r}\bH_{12}^{r*}+
\frac{P}{m_r}\bH^r\bH^{r*}+\bI_{m_1}\right)$, since
$\by_1 = \sqrt\frac{{P}}{m_2}\bH_{12}^r\bx_2 + \sqrt\frac{{P}}{m_r}\bH^r\bx_r
+ \bn_1$.
Similarly,
\begin{eqnarray}
\label{dmtdummy4}
I(\bx_r;\by_2|\bx_2) = \log\frac{L_{1r}^2}{L_{12}},
\end{eqnarray}
where $L_{1r}^2 = \det\left(\frac{P}{m_1}\bH_{12}\bH_{12}^{*}+
\frac{P}{m_r}\bG\bG^{*}+\bI_{m_2}\right)$.

To satisfy the compression rate constraints (\ref{compconstraintfd}), from
(\ref{dmtdummy1}), (\ref{dmtdummy2}), (\ref{dmtdummy3}), (\ref{dmtdummy4}), clearly
\begin{equation}
\label{leastnoisequantfd}
{\hat N} \ge
\frac{
\max\left\{\log \frac{L^{r2}_1}{L_{12}{\hat N}^{m_r}}, \log
\frac{L^{r1}_2}{L_{21}{\hat N}^{m_r}}\right\}
}
{
\min
\left\{
\log\frac{L_{1r}^2}{L_{12}}, \log \frac{L_{2r}^1}{L_{21}}
\right\}
}.
\end{equation}
We choose $\hat{N}$ to satisfy the equality (\ref{leastnoisequantfd}).
From \cite{Zheng2003},
to compute $d_{12}(r_{12}, r_{21})$, it is sufficient to find the negative of
the exponent of the $\SNR$ of outage probability at $T_2$,
where outage probability at $T_2$, $P_{out}(r_{12}\log \SNR)$, is defined as
\begin{eqnarray*}
P_{out}(r_{12}\log \SNR) &=& P(R_{12} \le r_{12}\log \SNR)
\end{eqnarray*}
From (\ref{cfratefd}, \ref{mis-rd}),
\begin{equation}
R_{12}  = \log \frac{L^{r2}_{1}}{({\hat N}+1)^{m_r}},
\end{equation}
where ${\hat N}$ is given in (\ref{leastnoisequantfd}).
Then,
\begin{eqnarray*}
P_{out}(r_{12}\log \SNR) &=& P\left(\log \frac{L^{r2}_{1}}{({\hat N}+1)^{m_r}} \le r_{12}\log \SNR\right),\\
&=& P\left(\frac{L^{r2}_{1}}{({\hat N}+1)^{m_r}} \le \SNR^{r_{12}}\right).
\end{eqnarray*}
Choose $l \in \bbZ$ such that $({\hat N}+1)^{m_r} \le l\left(\left(\frac{L^{r2}_1}{L_{1r}^2}\right)^{1/{m_r}} + 1\right)^{m_r}$, where ${\hat N}$ is such
that it meets the equality in (\ref{leastnoisequantfd}).
Then,
\begin{eqnarray}
\label{reduction1}
P_{out}(r_{12}\log \SNR) &\expl & P\left(\frac{L^{r2}_{1}}
{l\left(\left(\frac{L^{r2}_1}{L_{1r}^2}\right)^{1/{m_r}} + 1\right)^{m_r}} \le \SNR^{r_{12}}\right),\\\label{reduction2}
&=& P\left(\left(\frac{  (L^{r2}_{1})^{1/m_r} (L_{1r}^2)^{1/m_r}
}
{
l^{1/m_r}\left((L^{r2}_1)^{1/m_r} + (L_{1r}^2)^{1/m_r}\right)
}\right)^{m_r} \le \SNR^{r_{12}}\right),\\\label{reduction3}
&=& P\left(\frac{  (L^{r2}_{1})^{1/m_r} (L_{1r}^2)^{1/m_r}
}
{
(L^{r2}_1)^{1/m_r} + (L_{1r}^2)^{1/m_r}
} \le l^{1/m_r}\SNR^{r_{12}/{m_r}}\right),\\\label{reduction4}
&\expl& P\left(\frac{  (L^{r2}_{1})^{1/m_r} (L_{1r}^2)^{1/m_r}
}
{
(L^{r2}_1)^{1/m_r} + (L_{1r}^2)^{1/m_r}
} \le \SNR^{r_{12}/{m_r}}\right),
\end{eqnarray}
where the last equality follows because multiplying SNR by a constant does not change DM-tradeoff.
From here on we follow \cite{Yuksel2007} to compute the exponent of the
$P_{out}(r_{12}\log \SNR)$.
Let
\begin{equation}
\label{defnl1l2r}L^{r2}_{1l} =
\det\left(\frac{P}{m_1}\bH_{1}^{r2}\bH_{1}^{r2*} + \bI_{m_r+m_2}\right).
\end{equation}
Then clearly from (\ref{ents-rd}), $L^{r2}_{1l} \le L^{r2}_{1}$, therefore using
Lemma 2 \cite{Yuksel2007}, it follows that
\begin{eqnarray}
\label{reduction5}
P_{out}(r_{12}\log \SNR)&\le &
P\left(\frac{ (L^{r2}_{1l})^{1/m_r} (L_{1r}^2)^{1/m_r}
}
{
(L^{r2}_{1l})^{1/m_r} + (L_{1r}^2)^{1/m_r}
} \le \SNR^{r_{12}/{m_r}}\right).\end{eqnarray}

Moreover, notice that for non-negative random variables $X$ and $Y$ and a constant $c$ \cite{Yuksel2007},
$P(XY /(X+Y) < c) \le P(X < 2c) + P(Y <2c)$, thus,
\begin{eqnarray}
\label{reduction6}
P_{out}(r_{12}\log \SNR)&\le &
P\left( (L^{r2}_{1l})^{1/m_r} \le 2\SNR^{r_{12}/{m_r}} \right)   +
P\left((L_{1r}^2)^{1/m_r}  \le 2\SNR^{r_{12}/{m_r}} \right), \\\label{reduction7}
&\expeq &P\left( L^{r2}_{1l} \le \SNR^{r_{12}} \right)   +
P\left( L_{1r}^2 \le \SNR^{r_{12}} \right), \\ \nonumber
&\expeq & \SNR^{-d_1(r_{12})}  + \SNR^{-d_2(r_{12})},\\
&\expeq & \SNR^{-\min\left\{d_1(r_{12}), \ d_2(r_{12})\right\}}.
\end{eqnarray}
Therefore, to lower bound the DM-tradeoff we need to find out the
outage exponents $d_1(r_{12})$ and $d_2 (r_{12})$ of $L^{r2}_{1l}$
and $L_{1r}^2$. Notice that, however, $L^{r2}_{1l}$ is the mutual
information between $T_1$ and $T_2$ by choosing the covariance
matrix to be $\frac{P}{m_1}\bI_{m_1}$\footnote{$P$ taking the role
of $\SNR$.}, and allowing the relay and $T_2$ to cooperate perfectly.
From \cite{Zheng2003}, choice of $\frac{P}{m_1}\bI_{m_1}$ as the 
covariance matrix does not change the optimal 
DM-tradeoff, therefore,
$d_1(r_{12}) = (m_1-r_{12})(m_r+m_2-r_{12})$. Similar argument
holds for $L_{1r}^2$, by noting that $L_{1r}^2$ is the mutual
information between $T_1$ and $T_2$ if the relay and $T_1$ were
co-located and could cooperate perfectly, while using covariance
matrix $\frac{P}{m_1+m_r}\bI_{m_1+m_r}$. Thus, $d_2(r_{12}) =
(m_1+m_r-r_{12})(m_2-r_{12})$. Thus, for $T_1$ to $T_2$
communication, the achievable DM-tradeoff with CF strategy meets the
upper bound (Lemma \ref{dmtupboundfd}). A similar result can be
obtained for $T_2$ to $T_1$ communication by choosing an appropriate
$n \in \bbZ$ such that $({\hat N}+1)^{m_r} \le
n\left(\left(\frac{L^{r1}_2}{L_{2r}^1}\right)^{1/{m_r}} +
1\right)^{m_r}$, where ${\hat N}$ is such that it meets the equality
in (\ref{leastnoisequantfd}) and by carrying out the outage exponent
analysis of $R_{21} = \log \frac{L_2^{r1}} {({\hat N} +1)^{m_r}}$
and lower bounding $L_2^{r1}$ by $L_{2l}^{r1}$, where
 $L_{2l}^{r1} = \det\left(\frac{P}{m_2}\bH_{2}^{r1}\bH_{2}^{r1*} + \bI_{m_r+m_1}\right)$.
\end{proof}

\subsection{Half-Duplex Two-Way Relay Channel}
\label{subsec:dmthd}
In this section we compute the DM-tradeoff of the half-duplex two-way
relay channel where all the nodes ($T_1$, $T_2$ and the relay) are half-
duplex. For the half-duplex case, the achievable rate regions are protocol
dependent and the optimal protocol is unknown in general \cite{Kim2007,Kim2008,Boche2007}.
Here we compute the DM-tradeoff of a three phase protocol, that is intuitively
optimal (difficult to prove), where for $t_1$ fraction of the time slot
$T_1$ transmits to both $T_2$ and the relay, $t_2$ fraction of the time slot
$T_2$ transmits to $T_1$ and the relay, and for the rest $(1-t_1+t_2)$ fraction
of the time slot the relay transmits to both $T_1$ and $T_2$.

For this communication protocol the rates $R_{12}$ and $R_{21}$ are upper
bounded by the following expressions.
\begin{eqnarray*}
R_{12} \le\max_{t_1,t_2} \min\left\{t_1 I(\bx_1;\by_r,\by_2), t_1 I(\bx_1;\by_2) +
(1-t_1-t_2) I(\bx_r;\by_2) \right\}, \\
R_{21} \le \max_{t_1,t_2}\min\left\{t_2 I(\bx_2;\by_r,\by_1), t_2 I(\bx_2;\by_1)
 + (1-t_1-t_2) I(\bx_r;\by_1) \right\},
\end{eqnarray*}
where the first argument in the minimum is obtained by allowing the
relay and the $T_2$ ($T_1$) to collaborate in the receive mode, and the
second argument is obtained by simply adding the maximum
mutual information possible at $T_2$ ($T_1$) while in receiving mode.
Using the rate region expression, we define the upper bound on the
DM-tradeoff of the half-duplex two-way relay channel as follows.

From the definition of $L_{1l}^{r2}$ (\ref{defnl1l2r}),
\begin{eqnarray}\nonumber
P\left(t_1 I(\bx_1;\by_r,\by_2) \le r_{12}\log \SNR\right) & \expeq &
\label{upbounddmthd1}
P\left(t_1 \log L_1^{r2} \le r_{12}\log \SNR\right), \\
&\bydef& \SNR^{-d_{bc}^{12}(r_{12})},  \text{and } \\ \nonumber
P\left(t_1 I(\bx_1;\by_2) + (1-t_1-t_2) I(\bx_r;\by_2)\right) &\expeq& P\left(t_1 \log
L_{12} + (1-t_1-t_2) \log L_{r2} \le r_{12}\log \SNR\right), \\\label{upbounddmthd2}
&\bydef& \SNR^{-d_{mac}^{12}(r_{12})},
\end{eqnarray}
where $L_{2r} = \det\left(\bI_{m_2} + \frac{P}{m_r}\bG\bG^*\right)$.
Thus, $d_{12}(r_{12},r_{21}) \le \max_{t_1,t_2}\min\left\{d_{bc}^{12}(r_{12}), \ d_{mac}^{12}(r_{12})\right\}$. Similarly we can obtain upper bound for $d_{21}(r_{12},r_{21})$ by replacing $t_1$ by $t_2$ in (\ref{upbounddmthd1}, \ref{upbounddmthd2}).

To achieve this upper bound we consider the CF strategy of subsection
\ref{subsec:dmtfd},
except that in this case the compression signal ${\hat y}$ is chosen such that 
it is jointly typical with the received signals $y_{rt_{1}}$ and $y_{rt_{2}}$
 received in time
$t_1$ and $t_2$ from $T_1$ and $T_2$, respectively \footnote{In \cite{Kim2008} a
similar strategy has been proposed, but there, two
separate compression signals are chosen that are jointly typical with
$y_{rt_{1}}$ and $y_{rt_{2}}$ individually, and then a
deterministic function of the two compression signals is transmitted from the
relay, which results in a different rate region expression from the one obtained here.}.
With this CF strategy the achievable rate region is given by
\begin{eqnarray*}
R_{12} &\le& t_1 I(\bx_1;\by_2{\hat \by}|\bx_r,\bx_2)),\\
R_{21} &\le& t_2 I(\bx_2;\by_1{\hat \by}|\bx_r,\bx_1)),
\end{eqnarray*}
subject to the following compression rate constraint
\begin{eqnarray}
\label{compconstrainthd}
(t_1+t_2)\max\{I(\by_r; \hat{\by} | \bx_r \bx_1 \by_1), I(\by_r; \hat{\by} | \bx_r \bx_2 \by_2)\} \le
(1-(t_1+t_2))\min\{I(\bx_r;\by_1 | \bx_1), I(\bx_r;\by_2 | \bx_r \bx_2)\}.
\end{eqnarray}

To compute these rates, we let $\bx_1$, $\bx_2$ and $\bx_r$ to be the same as
in the full-duplex case and ${\hat \by} = \by_{rt_{1}}+\by_{rt_{2}}+\bn_q$,
where $\bn_q$ is the complex Gaussian vector with zero mean and covariance matrix ${\hat N}\bI_r$.
Following the same steps as in (\ref{leastnoisequantfd}) to (\ref{reduction7}),
we obtain
\begin{eqnarray}
\nonumber
P(R_{12}\le r_{12}\log \SNR) & \expl & P(t_1\log L_{1l}^{r2} \le r_{12}\log\SNR) +\\
\nonumber
&&
P\left(\frac{(2(t_1+t_2)-1)t_1}{t_1+t_2}\log L_{12} + \frac{(1-(t_1+t_2))t_1}{t_1+t_2}\log L_{1r}^2  \le r_{12}\log\SNR\right), \\ \nonumber
& \bydef & \SNR^{-d_{bc}^{12}(r_{12})} + \SNR^{-d^{12'}_{mac}(r_{12})}, \\
\label{lowerbddmthd}
& \expeq &
\SNR^{
-\min\left\{
d_{bc}^{12}(r_{12}),
d^{12'}_{mac}(r_{12})
\right\}
}.
\end{eqnarray}
Thus the achievable $d_{12}(r_{12},r_{21}) \le \max_{t_1,t_2}
\min\left\{d_{bc}^{12}(r_{12}), d^{12'}_{mac}(r_{12})\right\}$.
Note that the expression for $d_{12}(r_{12},r_{21})$ is independent of
$r_{21}$, and because of symmetry in $R_{12}$ and $R_{21}$ expressions,
similar bounds can be obtained for $R_{21}$ by replacing $t_1$ with $t_2$, and
is given by
\begin{eqnarray*}
P(R_{21}\le r_{21}\log \SNR) & \le & P(t_1\log L_{2l}^{r1} \le r_{21}\log\SNR) +\\
&&
P\left(\frac{(2(t_1+t_2)-1)t_2}{t_1+t_2}\log L_{12} + \frac{(1-(t_1+t_2))t_2}{t_1+t_2}\log L_{2r}^1  \le r_{21}\log\SNR\right), \\
& \bydef & \SNR^{-d_{bc}^{21}(r_{21})} + \SNR^{-d^{21'}_{mac}(r_{21})},
\end{eqnarray*}
which implies \begin{eqnarray}
\label{lowerbddmthd2}
d_{21}(r_{12},r_{21}) \le \max_{t_1,t_2}
\min\left\{d_{bc}^{21}(r_{21}), d^{21'}_{mac}(r_{21})\right\}.\end{eqnarray}
It is clear that the lower bound (\ref{lowerbddmthd}, \ref{lowerbddmthd2}) and the upper bound (\ref{upbounddmthd1},\ref{upbounddmthd2}) on the DMT of the half-duplex two-way relay channel do
not match for the general case.
By comparing the achievable DM-tradeoff and the upper bound, the next Theorem
characterizes the cases for which CF strategy is optimal.
\begin{thm}
The proposed CF strategy achieves the optimal DM-tradeoff of the half-duplex
two way relay channel if
\begin{itemize}
\item the bottleneck of the channel is the broadcast cut, i.e.
$d_{bc}^{12}(r_{12}) \le d^{12'}_{mac}(r_{12})$ and
correspondingly in the upper bound
$d_{bc}^{12}(r_{12}) \le d^{12}_{mac}(r_{12})$, and with similar relation for
$d_{bc}^{21}(r_{21}) $ and $d_{mac}^{21}(r_{21}) $also.
\item otherwise if
$\frac{(2(t_1+t_2)-1)t_1}{t_1+t_2}\log L_{12} + \frac{(1-(t_1+t_2))t_1}{t_1+t_2}\log L_{1r}^2 = t_1 \log L_{12} + (1-t_1-t_2) L_{r2}$,
and with similar relation for $T_2$ to $T_1$ communication.
\end{itemize}
\end{thm}
\begin{proof} Follows immediately by comparing the lower bound
(\ref{lowerbddmthd}) and the upper bound (\ref{upbounddmthd1},\ref{upbounddmthd2}) on the DM-tradeoff.
\end{proof}

{\bf Discussion:}
In this section we showed that the CF strategy achieves the optimal DM-tradeoff
of the two-way relay channel for the full-duplex case, in general, and for the
half-duplex case in some cases. For both the full-duplex and half-duplex
case we upper bounded the DM-tradeoff allowing different nodes to collaborate
 with each other while transmitting or receiving. For the full-duplex case,
we modified the CF strategy of \cite{Cover1979} \footnote{The same strategy can also
be found in \cite{Kim2008}} and showed that it decouples the two-way relay channel
into two one-way relay channel and achieves optimal DM-tradeoff on each
of the two one-way relay channels. For the half-duplex case, as observed
before, the achievable rate region and consequently the DM-tradeoff depends
on the communication protocol. We used a three phase protocol that makes use of
all the direct links between $T_1$, $T_2$, and the relay.
For the three phase protocol we proposed a modified CF
 strategy where the compression signal is chosen such that it is jointly 
typical with the signals received at the relay node in phase $1$ and $2$.
Using this CF strategy, we obtained a lower bound on the DM-tradeoff that is
shown to match with the upper bound under some conditions. For the general
case also, we believe that the proposed CF should be optimal in terms of
achieving the DM-tradeoff, however, showing that is quite difficult because
of the different mutual information quantities involved as well as the maximization over the time durations of phase $1$ and $2$.

Our result for the full-duplex case is similar to \cite{Yuksel2007},
where it is shown that the CF strategy achieves the optimal DM-tradeoff in
one-way relay channel. For the half-duplex case, however, because of three phase
communication protocol and added compression rate constraints we are
unable to reach the same conclusion of \cite{Yuksel2007} in general,
that CF achieves the optimal DM-tradeoff in half-duplex one-way relay channel.

\section{Conclusion}
\label{conc}
In the first part of the paper,
we addressed the problem of finding optimal relay beamformers to maximize the
achievable rate region of the two-way relay channel with multiple relays,
when each relay uses AF.
The use of AF strategy is motivated by the fact that
 all the other known relay strategies such as DF, partial DF
and CF, do not work well in the presence of multiple relays, and
moreover, AF is quite simple to implement.

For the case when both the terminals $T_1$ and $T_2$ have a single antenna
and each relay has an arbitrary number of antennas,
we found an iterative algorithm to compute the optimal relay beamformers.
The algorithm is equivalent to solving a
power minimization problem subject to SINR constraints at each step.
The power minimization problem at each step is non-convex,
however, for which it is sufficient to satisfy the KKT conditions to obtain the
optimal solution.

The derived optimal AF strategy maximizes the rate region with AF, but
is restricted to the case of a single antenna at $T_1$ and $T_2$, and
cannot be extended easily for the multi-antenna case. Moreover, it also
requires each relay to have global CSI, and does not have a closed form achievable rate region
expression. To relax the single antenna restriction and global CSI requirement,
 we then proposed a dual channel matching strategy, which requires local CSI,
 and showed that the gap between the rate region of the optimal AF and dual channel matching is quite small when both $T_1$ and $T_2$ have a single antenna.
The dual channel matching works for any number of antennas at $T_1$ and $T_2$,
and has a closed form expression for the achievable rate region. We then
compared the achievable rate region of the dual channel matching with an
upper bound to quantify the loss while using dual channel matching.
The analytical expressions of the lower and the upper bound did not match, and
we used simulations to show that the gap is quite small.
In the asymptotic regime of $K \rightarrow \infty$, however,
using the analytical expressions, we proved that the achievable rate region
of the dual channel matching, is only a constant term away from the upper bound. Thus, we obtained the capacity scaling law for the two-way relay channel.
Compared to \cite{Gastpar2002b, Bolcskei2006}, our capacity scaling law for the
two-way relay channel shows that with two-way relay channel,
there is a two-fold increase in the capacity compared to
unidirectional communication.

In the second part of the paper, we considered the problem of finding
coding strategies that achieve the
optimal DM-tradeoff in a two-way relay channel with a single relay node, in the
presence of direct path between $T_1$ and $T_2$. We
showed that the CF strategy achieves the optimal DM-tradeoff of the
full-duplex two-way relay channel, by first
decoupling the two-way relay channel into two one-way relay channels, and
achieving the optimal DM-tradeoff on each of the two one-way relay channel.
For the half-duplex case we showed that a modified CF strategy for
a three phase transmission protocol achieves the optimal DM-tradeoff for some
cases.

\section{Acknowledgments}
We are thankful to Harish Ganapathy for pointing out reference \cite{Wiesel2006}.


\begin{thebibliography}{10}
\bibitem{Muelen1971}
E.~Van~der Meulen, ``Three terminal communication channels,'' \emph{Adv. Appl.
  Probab.}, vol.~3, pp. 120--154, 1971.

\bibitem{Rankov2005}
B.~Rankov and A.~Wittneben, ``Spectral efficient signaling for half-duplex
  relay channels,'' in \emph{Asilomar Conference on Signals, Systems, and
  Computers, Pacific Grove, CA}, Oct.-Nov.2005 2005, pp. 1066--1071.

\bibitem{Kim2007}
S.~Kim, P.~Mitran, and V.~Tarokh, ``Performance bounds for bi-directional coded
  cooperation protocols,'' \emph{{IEEE} Trans. Inf. Theory}, submitted Mar.
  2007, available on http://arxiv.org/abs/cs/0703017.

\bibitem{Kim2008}
S.~Kim, P.~Mitran, and V.~Tarokh, ``Achievable rate regions for bi-directional relaying,'' \emph{{IEEE}
  Trans. Inf. Theory}, submitted Aug. 2008, available on 
  {http://arxiv.org/PS\_cache/arxiv/pdf/0808/0808.0954v1.pdf}


\bibitem{Boche2007}
T.~Oechtering and H.~Boche, ``Optimal tranmsit strategies in multi-antenna
  bidirectional relaying,'' in \emph{IEEE Intern. Conf. on Acoustics, Speech,
  and Signal Processing (ICASSP '07), Honolulu, Hawaii, USA}, vol.~3, April
  2007, pp. 145--148.

\bibitem{Dina2006}
S.~Katti, R.~Hariharan, W.~Hu, D.~Katabi, M.~Medard, and J.~Crowcroft, ``Xors
  in the air: Practical wireless network coding,'' \emph{ACM SIGCOMM Pisa,
  Italy, Sept. 2006}, pp. 243--254.

\bibitem{Yeung2003}
S.-Y. Li, R.~Yeung, and N.~Cai, ``Linear network coding,'' \emph{{IEEE} Trans.
  Inf. Theory}, vol.~49, no.~2, pp. 371--381, Feb. 2003.

\bibitem{Rankov2006}
B.~Rankov and A.~Wittneben, ``Achievable rate regions for the two-way relay
  channel,'' in \emph{IEEE Int. Symposium on Information Theory (ISIT),
  Seattle, USA}, July 2006 2006, pp. 1668--1672.

\bibitem{Tse2008}

A.~Avestimehr, A.~Sezgin, and D.~Tse, ``Approximate capacity of the two-way
  relay channel: A deterministic approach,'' in \emph{Allerton Conference on
  Communication, Control, and Computing,, Monticello, IL}, 2008, available on.
 {http://arxiv.org/PS\_cache/arxiv/pdf/0808/0808.3145v1.pdf}


\bibitem{Nazer2007}
B.~Nazer and M.~Gastpar, ``Computation over multiple-access channels,''
  \emph{{IEEE} Trans. Inf. Theory}, vol.~53, no.~10, pp. 3498--3516, Oct. 2007.

\bibitem{Narayan2007}
M.~P. Wilson, K.~Narayanan, H.~Pfister, and A.~Sprintson, ``Joint physical
  layer coding and network coding for bi-directional relaying,'' in
  \emph{Allerton Conference on Communication, Control, and Computing,
  Monticello, IL}, 2007.

\bibitem{Popovski2007}
P.~Popovski and H.~Yomo, ``Physical network coding in two-way wireless relay
  channels,'' in \emph{IEEE International Conference on Communications, 2007.
  ICC '07.}, 24-28 June 2007, pp. 707--712.

\bibitem{Shengli2008}
Z.~Shengli and S.~Liew, ``The capacity of two way relay channel,'' available
  on. {http://arxiv.org/ftp/arxiv/papers/0804/0804.3120.pdf}


\bibitem{Kramer2003}
G.~Kramer and S.~Savari, ``On networks of two-way channels,'' in
  \emph{Algebraic Coding Theory and Information Theory, DIMACS Workshop,
  Rutgers University, DIMACS Series in Discrete Mathematics and Theoretical
  Computer Science}, vol.~68, Dec 2003, {available on}
  http://cm.bell-labs.com/who/gkr/, pp. 133--143.

\bibitem{Kramer2005}
G.~Kramer, M.~Gastpar, and P.~Gupta, ``Cooperative strategies and capacity
  theorems for relay networks,'' \emph{{IEEE} Trans. Inf. Theory}, vol.~51,
  no.~9, pp. 3037--3063, Sept. 2005.

\bibitem{Yi2007}
Z.~Yi and I.-M. Kim, ``Joint optimization of relay-precoders and decoders with
  partial channel side information in cooperative networks,'' \emph{{IEEE} J.
  Sel. Areas Commun.}, vol.~25, no.~2, pp. 447--458, February 2007.

\bibitem{Boyd2004}
S.~Boyd and L.~Vandenberghe, \emph{Convex Optimization}.\hskip 1em plus 0.5em
  minus 0.4em\relax Cambridge University Press, 2004.

\bibitem{Wiesel2006}
A.~Wiesel, Y.~Eldar, and S.~Shamai, ``Linear precoding via conic optimization
  for fixed {MIMO} receivers,'' \emph{{IEEE} Trans. Signal Process.}, vol.~54,
  no.~1, pp. 161--176, Jan. 2006.

\bibitem{Cover2004}
T.~Cover and J.~Thomas, \emph{Elements of Information Theory}.\hskip 1em plus
  0.5em minus 0.4em\relax John Wiley and Sons, 2004.

\bibitem{Bolcskei2006}
H.~Bolcskei, R.~Nabar, O.~Oyman, and A.~Paulraj, ``Capacity scaling laws in
  {MIMO} relay networks,'' \emph{{IEEE} Trans. Wireless Commun.}, vol.~5,
  no.~6, pp. 1433--1444, June 2006.

\bibitem{Zheng2003}
L.~Zheng and D.~Tse, ``Diversity and multiplexing: A fundamental tradeoff in
  multiple-antenna channels,'' \emph{{IEEE} Trans. Inf. Theory}, vol.~49,
  no.~5, pp. 1073--1096, May 2003.

\bibitem{Azarian2005}
K.~Azarian, H.~El~Gamal, and P.~Schniter, ``On the achievable
  diversity-multiplexing tradeoff in half-duplex cooperative channels,''
  \emph{{IEEE} Trans. Inf. Theory}, vol.~51, no.~12, pp. 4152--4172, Dec. 2005.

\bibitem{Belfiore2007}
C.~Yang and J.-C. Belfiore, ``Optimal space time codes for the {MIMO}
  amplify-and-forward cooperative channel,'' \emph{{IEEE} Trans. Inf. Theory},
  vol.~53, no.~2, pp. 647--663, Feb. 2007.

\bibitem{EliaDec72005}
P.~Elia and P.~Vijay~Kumar, ``Approximately universal optimality over several
  dynamic and non-dynamic cooperative diversity schemes for wireless
  networks,'' \emph{available at http://arxiv.org/pdf/cs.it/0512028}, Dec 7,
  2005.

\bibitem{Yuksel2007}
M.~Yuksel and E.~Erkip, ``Multiple-antenna cooperative wireless systems: A
  diversity-multiplexing tradeoff perspective,'' \emph{{IEEE} Trans. Inf.
  Theory}, vol.~53, no.~10, pp. 3371--3393, Oct. 2007.

\bibitem{Mitran2008}
P.~Mitran, ``The diversity-multiplexing tradeoff for independent parallel mimo
  channels,'' in \emph{IEEE International Symposium on Information Theory,
  Toronto, 2008}, July 2008, pp. 2366--2370.

\bibitem{Cover1979}
T.~Cover and A.~El~Gamal, ``Capacity theorems for relay channels,''
  \emph{{IEEE} Trans. Inf. Theory}, vol.~25, no.~5, pp. 572--584, Sept. 1979.

\bibitem{Cioffi2006}
M.~Mohseni, R.~Zhang, and J.~Cioffi, ``Optimized transmission for fading
  multiple-access and broadcast channels with multiple antennas,'' vol.~24,
  no.~8, pp. 1627--1639, Aug. 2006.

\bibitem{Munoz-Medina2007}
O.~Munoz-Medina, J.~Vidal, and A.~Agustin, ``Linear transceiver design in
  nonregenerative relays with channel state information,'' \emph{{IEEE} Trans.
  Signal Process.}, vol.~55, no.~6, pp. 2593--2604, June 2007.

\bibitem{Dana2006}
A.~Dana and B.~Hassibi, ``On the power efficiency of sensory and ad hoc
  wireless networks,'' \emph{{IEEE} Trans. Inf. Theory}, vol.~52, no.~7, pp.
  2890--2914, July 2006.

\bibitem{Telatar1999}
E.~Telatar, ``Capacity of multi-antenna gaussian channels,'' \emph{European
  Trans. on Telecommunications}, vol.~10, no.~6, pp. 585--595, Nov./Dec. 1999.

\bibitem{Gastpar2002b}
M.~Gastpar and M.~Vetterli, ``On the capacity of wireless networks: the relay
  case,'' in \emph{INFOCOM 2002. Twenty-First Annual Joint Conference of the
  IEEE Computer and Communications Societies. Proceedings. IEEE}, vol.~3, 23-27
  June 2002, pp. 1577--1586vol.3.

\bibitem{Hassibi2003}
B.~Hassibi and B.~Hochwald, ``How much training is needed in multiple-antenna
  wireless links?'' \emph{{IEEE} Trans. Inf. Theory}, vol.~49, no.~4, pp.
  951--963, April 2003.

\bibitem{Biglieri2001}
E.~Biglieri, G.~Caire, and G.~Taricco, ``Limiting performance of block-fading
  channels with multiple antennas,'' \emph{Information Theory, IEEE
  Transactions on}, vol.~47, no.~4, pp. 1273--1289, May 2001.


\end{thebibliography}
\end{document}